\documentclass{nature}
\usepackage{comment}
\usepackage{cite}
\usepackage[T1]{fontenc}
\usepackage[utf8]{inputenc}
\usepackage{graphicx}
\usepackage{float}
\usepackage{caption}
\usepackage{amssymb}
\usepackage{amsmath}
\usepackage{mathtools}
\usepackage{dcolumn}
\usepackage{pdfpages}
\usepackage{color}
\usepackage{bm}
\usepackage[normalem]{ulem}
\usepackage[T1]{fontenc}
\usepackage[utf8]{inputenc}
\usepackage{multibib}
\usepackage[section]{placeins}
\usepackage[labelfont=bf]{caption}
\usepackage{enumitem}  
\usepackage{chemformula}
\usepackage{upgreek}

\newcommand{\beq}{\begin{equation}}
\newcommand{\eeq}{\end{equation}}
\newcommand{\beqn}{\begin{eqnarray}}
\newcommand{\eeqn}{\end{eqnarray}}

\begin{document}

\title{Selective braiding of different anyons in the even-denominator fractional quantum Hall effect}

\author{Jehyun Kim$^{1,}$\footnote{These authors contributed equally to this work.}, Amit Shaer$^{1,*}$, Ravi Kumar$^1$, Alexey Ilin$^1$, Kenji Watanabe$^{2}$, Takashi Taniguchi$^{3}$, Ady Stern$^1$, David F. Mross$^1$ and Yuval Ronen$^{1}$\footnote{Corresponding author: yuval.ronen@weizmann.ac.il}}

\maketitle

\begin{affiliations}

\item Department of Condensed Matter Physics, Weizmann Institute of Science, Rehovot 76100, Israel.

\item Research Center for Functional Materials, National Institute for Materials Science, 1-1 Namiki, Tsukuba 305-0044, Japan.
\item International Center for Materials Nanoarchitectonics, National Institute for Materials Science, 1-1 Namiki, Tsukuba 305-0044, Japan.
\\

\end{affiliations}

\begin{abstract}
Even-denominator quantum Hall states can host several types of anyons with distinct exchange statistics. Depending on the anyon type, exchanging two quasiparticles can impart a phase to the many-body wave function or even transform it into a different state. Here, we realize a gate-tunable Fabry-P\'erot interferometer with an embedded antidot that provides local control over the number of anyons within the interference loop. By independently tuning the magnetic field, carrier densities across the device, and the antidot potential, we access regimes in which localized anyons form reproducibly and measure the associated statistical phases $e^{i \theta_\mathrm{braid}}$. We resolve braiding phases of $\theta_{\mathrm{braid}}=\pi$ and $\theta_{\mathrm{braid}}=\frac{\pi}{2}$, which we attribute to $e/2$ quasiparticles encircling either $e/2$ or $e/4$ quasiparticles, respectively. We further observe switching between different anyon occupancies of the antidot over time, directly resolving individual anyon tunnelling events into the interference loop. Similar behavior occurs at filling factor one third. Our work addresses one of the two key challenges in observing non-Abelian braiding, which requires control of both localized and interfering anyon types.

\end{abstract}

\clearpage
\noindent\textbf{Introduction}

Emergent quasiparticles with fractional exchange statistics, known as anyons, occur naturally in fractional quantum Hall (FQH) states.\cite{Wilczek1982, feldman2021fractional} When an Abelian anyon encircles another, the many-body wave function changes by a non-trivial phase,\cite{Leinaas1977} whereas it remains unchanged for bosons and fermions. The consequences of braiding are more pronounced for non-Abelian anyons, where the exchange operation can rotate the wave function into an orthogonal state.\cite{nayak2008non,stern2008anyons} Even-denominator FQH states are expected to host both types of anyons. Specifically, at half-filling, the $e/2$ quasiparticles are Abelian, while the $e/4$ may be non-Abelian, depending on the specific topological order.\cite{HalperinQH83,moore1991nonabelions,Greiter92} Both types of anyons could be probed in spatial\cite{nakamura2020direct,nakamura2023fabry,kim2024aharonov,werkmeister2025anyon,samuelson2024anyonic,ghosh2025anyonic} or time-domain\cite{bartolomei2020fractional,lee2023partitioning,glidic2023cross} interference experiments. Here, we focus on the former, where a beam of anyons is partitioned at quantum point contacts (QPCs) and guided around an FQH bulk containing localized anyons.

Braiding of Abelian anyons has been demonstrated at several filling factors, material platforms, and interferometer architectures. Experiments with Fabry-P\'erot interferometers (FPIs) have observed statistical phase jumps at one-third filling in GaAs,\cite{nakamura2020direct}monolayer graphene,\cite{werkmeister2025anyon, samuelson2024anyonic} and bilayer graphene.\cite{kim2024aharonov, kim2026aharonov} Anyon braiding in higher particle-like Jain states has so far been measured only in GaAs, using FPIs and optical Mach-Zehnder interferometers (OMZIs).\cite{nakamura2023fabry,ghosh2025anyonic} In these experiments, the charge of the interfering quasiparticles corresponded to the smallest one permitted by the FQH bulk. In contrast, experiments in hole-conjugate and even-denominator states reported interference of anyons whose charge followed the filling factor.\cite{ghosh2025coherent,kim2026aharonov} For example, the interfering charge was doubled to $2\times e/3$ at two-thirds and to $2\times e/4$ at half filling. Individual phase slips were not observed at these fillings despite a clear Aharonov-Bohm (AB) effect, which was attributed to fluctuations in the number of localized anyons. These fluctuations require careful analysis to extract braiding information about the bulk anyons (cf.~SI21 of Ref.~\citeonline{kim2026aharonov}) and prevent selective interference between different anyon types.

In this work, we control the localized bulk anyons and measure their statistical contribution by embedding a gate-defined antidot in a bilayer-graphene-based Fabry-P\'erot interferometer. At filling $\nu=-\frac{1}{2}$ we observe an interfering quasiparticle charge of $e/2$ via the AB effect, as previously reported.\cite{kim2026aharonov} When the antidot filling is weakly detuned from the FPI bulk filling, the interference pattern exhibits slow dynamics similar to those previously observed at $\nu=\frac{1}{3}$.\cite{werkmeister2025anyon, samuelson2024anyonic} A mechanism that leads to such dynamics is discussed in Ref.~\citeonline{mross2026}. Fluctuations occurring at random intervals on timescales of seconds to minutes shift the entire interference pattern by $\pi$, as expected when $e/2$ quasiparticles enter or leave the interference loop. These fluctuations are suppressed when the antidot filling is increased, until they reappear for antidot fillings close to $\nu_\mathrm{AD}=0$. There, $\pi/2$ phase jumps occur in addition to $\pi$ jumps, indicating the presence of two distinct processes: (i) braiding between two Abelian $e/2$ anyons; (ii) braiding between one Abelian $e/2$ and a localized $e/4$ anyon, expected to be non-Abelian.

\begin{figure}[H]
  \centering
  \includegraphics[width=.8\textwidth]{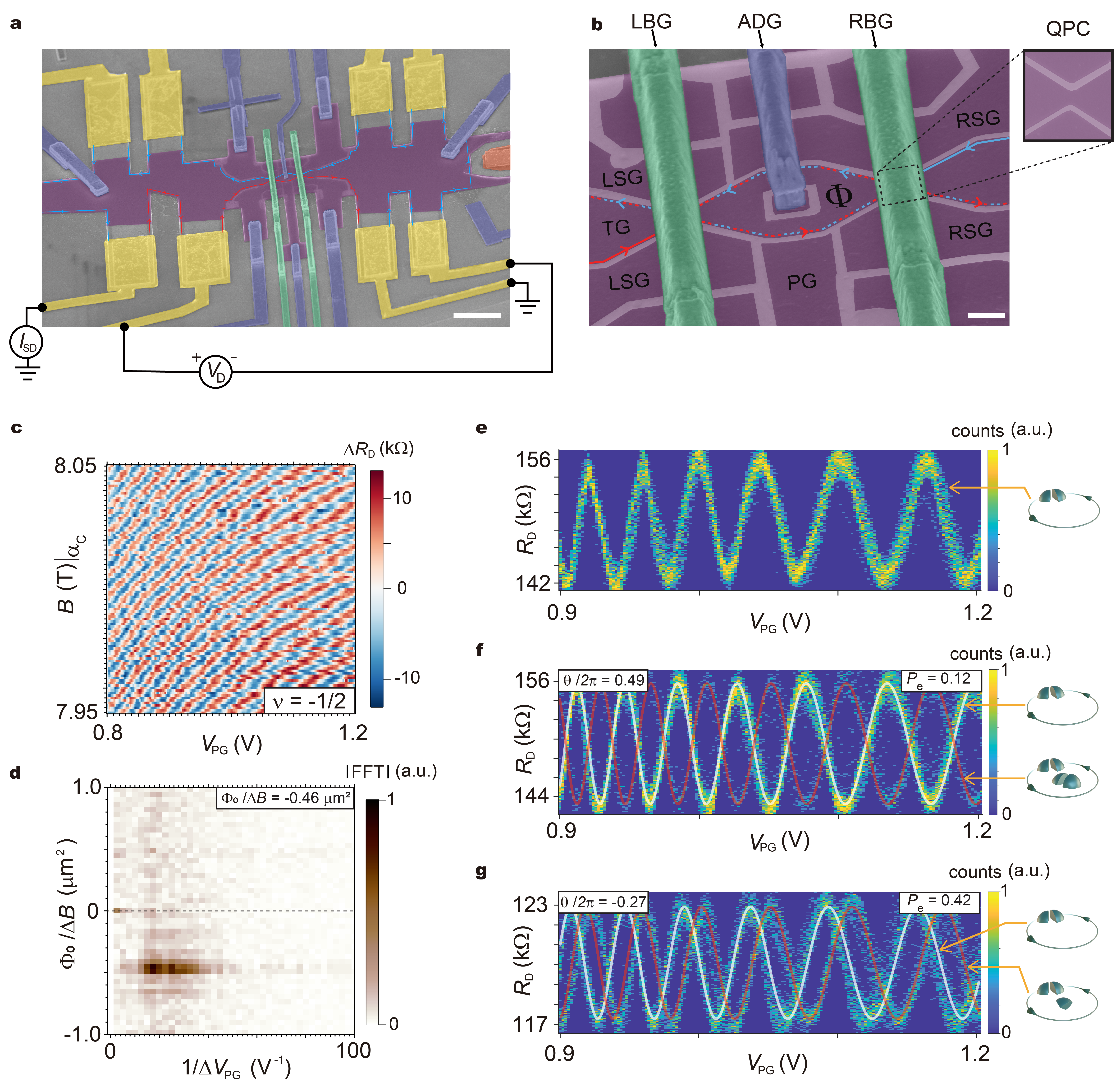}
  \caption{\textbf{Fabry-P\'erot interferometer with an embedded, gate-defined antidot.} (\textbf{a}) False-color scanning electron microscopy image of the bilayer graphene-based Fabry--P\'erot interferometer with an embedded antidot. A current $I_\mathrm{SD}$ enters through an ohmic contact (yellow), propagates along edge modes, and is partially transmitted through two quantum point contacts (QPCs) formed by the left and right split gates (LSG and RSG). The diagonal resistance is measured as $R_\mathrm{D} = (V_D^{+} - V_D^{-})/I_\mathrm{SD}$ in a single-ground configuration. Scale bar: $3~\upmu \mathrm{m}$. (\textbf{b}) Magnified view of the interference region near the antidot gate (ADG), defined by a central graphite island of area $0.1~\upmu \mathrm{m^2}$ within the lithographically defined interference area of approximately $1~\upmu \mathrm{m^2}$. The left and right air bridges (LBG and RBG), suspended approximately $200~\mathrm{nm}$ above the QPC regions, locally tune the electrostatic potential. Scale bar: $0.3~\upmu \mathrm{m}$. The QPC architecture with a $100~\mathrm{nm}$ gap is shown in the upper-right panel. (\textbf{c}) $\Delta$$R_\mathrm{D}$ at $\nu=-\frac{1}{2}$ displayed in a $B$-$V_\text{PG}$ plane at constant filling ($\alpha_c$), showing clear AB oscillations.  (\textbf{d}) Corresponding 2D-FFT analysis. (\textbf{e-g}) Histograms of $R_\mathrm{D}$ at $\nu = -\frac{1}{2}$ as a function of $V_\mathrm{PG}$, constructed from repeated $V_\mathrm{PG}$ sweeps under three different antidot gate voltages $V_\mathrm{ADG}$. The bin size was $50-100~\Omega$, determined by using 150 bins across the interference range. (e) $V_\mathrm{ADG}=0.4~\mathrm{V}$,  (f) $V_\mathrm{ADG}=0.235~\mathrm{V}$, (g) $V_\mathrm{ADG}=0.487~\mathrm{V}$. The white and red curves are cosine fits to the histogram peaks, demonstrating phase shifts of approximately $\pi$ in (f) and $\frac{\pi}{2}$ in (g).
}
  \label{fig1}
\end{figure}

\noindent\textbf{Fabry-P\'erot interferometer with an embedded, gate-defined antidot}

Our experiments are performed on an FPI device constructed on a high-mobility van der Waals heterostructure, following prior work in the field.\cite{deprez2021tunable,ronen2021aharonov,fu2023aharonov,zhao2022graphene} Bilayer graphene serves as the active two-dimensional layer and is encapsulated between hexagonal boron nitride dielectric layers.\cite{li2017even, zibrov2017tunable} The embedded antidot is defined by a graphite-gated region of area $ 0.1~\upmu\mathrm{m^2}$ within an FPI area of $A=1~\upmu\mathrm{m^2}$. A single continuous graphite gate controls the filling factors in the reservoirs and the bulk of the FPI. The heterostructure design and nanofabrication follow those described in our previous study.\cite{kim2024aharonov} Measurements are performed in perpendicular magnetic fields up to $B=12~\mathrm{T}$ and at a base temperature of $T=10~\mathrm{mK}$.

Figure~\ref{fig1}a shows a false-color scanning electron microscopy image of the device, with a magnified view of the interference region in Fig.~\ref{fig1}b. The top graphite layer is divided into seven regions by 40-nm-wide etched trenches, each of which is electrostatically connected via an air bridge. The saddle-point potentials of the QPCs are formed by tuning the left and right split gates, LSG and RSG, which guide the counter-propagating edge modes on opposite sides into close proximity and introduce tunneling between them (see SI1). Two air bridges, denoted LBG and RBG, provide additional control over tunneling. The area enclosed by the interfering quantum Hall edge modes is tuned via the plunger gate (PG), while the graphite antidot gate (ADG) sets the filling at the central island. 

We inject a bias current $I_\mathrm{SD}$ through the indicated contact, which propagates along the FQH edge. The current is collected at a single ground contact on the opposite side of the FPI, while the diagonal voltage $V_D$ is measured using a standard lock-in technique. In the weak backscattering limit, the diagonal resistance $R_\mathrm{D} = V_\mathrm{D}/I_\mathrm{SD}$ contains an oscillatory term $\Delta R_\mathrm{D} \propto \cos \theta$, where the interference phase $\theta$ includes the AB contribution and a braiding term
\begin{align}
    \theta = 2\pi\frac{e^*}{e} \frac{AB}{\Phi_0} + \sum_{\substack{
\mathrm{localized}\\\mathrm{anyon\ type }\ i }}N_i \theta^i_\mathrm{braid}~.
\label{eqn.theta}
\end{align}
Here, $e^*$ is the charge of the interfering anyon and $\theta^i_\mathrm{braid}$ its braiding phase acquired when encircling an anyon of type $i$ and $\Phi_0$ is the flux quantum.

Figure~\ref{fig1}c shows $\Delta$$R_\mathrm{D}$ as a function of the PG voltage, $V_\mathrm{PG}$, along the constant filling line $\alpha_c$, where a filling factor of $\nu=-\frac{1}{2}$ is maintained (see SI2). During this measurement, TG and ADG were kept at equal potentials, $V_\mathrm{TG}=V_\mathrm{ADG}$, such that the antidot is inactive. The two-dimensional fast Fourier transform (2D-FFT) in Fig.~\ref{fig1}d shows a single peak that is sharp in $1/\Delta B$, but broadened in $1/\Delta V_\mathrm{PG}$ because the capacitance between the interfering edge and the plunger gate varies with $V_\mathrm{PG}$. From the peak position $\Phi_0/\Delta B=-0.46$ $\mu\mathrm{m}^2$, we infer a magnetic-flux periodicity $\Delta \Phi \approx 2 \Phi_0$ as was previously observed,\cite{kim2026aharonov} confirming an interfering charge of $e/2$ in an FPI with a different QPC design (see SI3). 

Figures~\ref{fig1}e-g show $R_\mathrm{D}$ as a function  $V_\mathrm{PG}$ for three different antidot settings. As the measured resistance is not constant over time, we represent 100 $V_\mathrm{PG}$ sweeps recorded over a total of 9000 seconds as histograms of the $R_\mathrm{D}$ values measured for each $V_\mathrm{PG}$. 
The data taken at $\nu_\mathrm{AD}\approx0$ (Fig.~\ref{fig1}e) are representative of the interval $-\frac{1}{2} \lesssim \nu_\mathrm{AD} \lesssim 0$. In this regime, the histogram at each $V_\mathrm{PG}$ contains a single peak, indicating that fluctuations in the number of localized anyons are negligible. Figure~\ref{fig1}f was recorded for an antidot filling of $\nu_\mathrm{AD}\approx -0.5$, where the histogram contains a second peak that is shifted by $\pi$ with respect to the dominant one. This finding implies substantial fluctuations in the number of localized $e/2$ anyons with a time scale that is shorter than the sweep time, but longer than the time needed for a measurement of an individual $R_D$ value. Depending on the number of localized anyons, one of the two interference processes illustrated in the figure occurs.  Figure~\ref{fig1}g was recorded for $\nu_\mathrm{AD} \approx 0$  and shows two contributions of nearly equal strength, shifted by $\pi/2$. 

Phase jumps of $\pi$ and $\pi/2$ are theoretically expected when an interfering $e/2$ anyon encircles an interference area in which the number of localized $e/2$ or $e/4$ anyons changes by one, respectively; see Refs.~\citeonline{sarma2005,rosenow2012} and SI10.  In fact, the expected topological orders at the half-filled Landau level states allow for 
two topologically distinct $e/2$ anyons, characterized by even or odd fermion parity. The braiding phases of these $e/2$ anyons with $e/4$ anyons differ by $\pi$. If both types of $e/2$ anyons contributed equally to the interference, their oscillatory contributions to $R_\mathrm{D}$ would cancel one another when there are an odd number of localized $e/4$ anyons. Our observation of AB interference patterns that shift rigidly by $\frac{\pi}{2}$ implies that such a cancellation does not occur in our experiment, and
$e/2$ anyons with one parity dominate the interference process. We expect that the $e/2$ anyon with even-parity of fermions is the dominant one, since the tunneling of an odd-parity anyon involves exciting a neutral edge mode, giving it a larger scaling dimension\cite{yang2013} and making it more susceptible to dephasing.\cite{Bishara2008} 
The combined observations of Figs.~\ref{fig1}e-g show that applying a gate voltage to the antidot provides experimental selectivity of the localized anyon type around which braiding occurs.

\noindent\textbf{Antidot-tuned interference in integer and fractional quantum Hall states}

\begin{figure}[H]
  \centering
  \includegraphics[width=.95\textwidth]{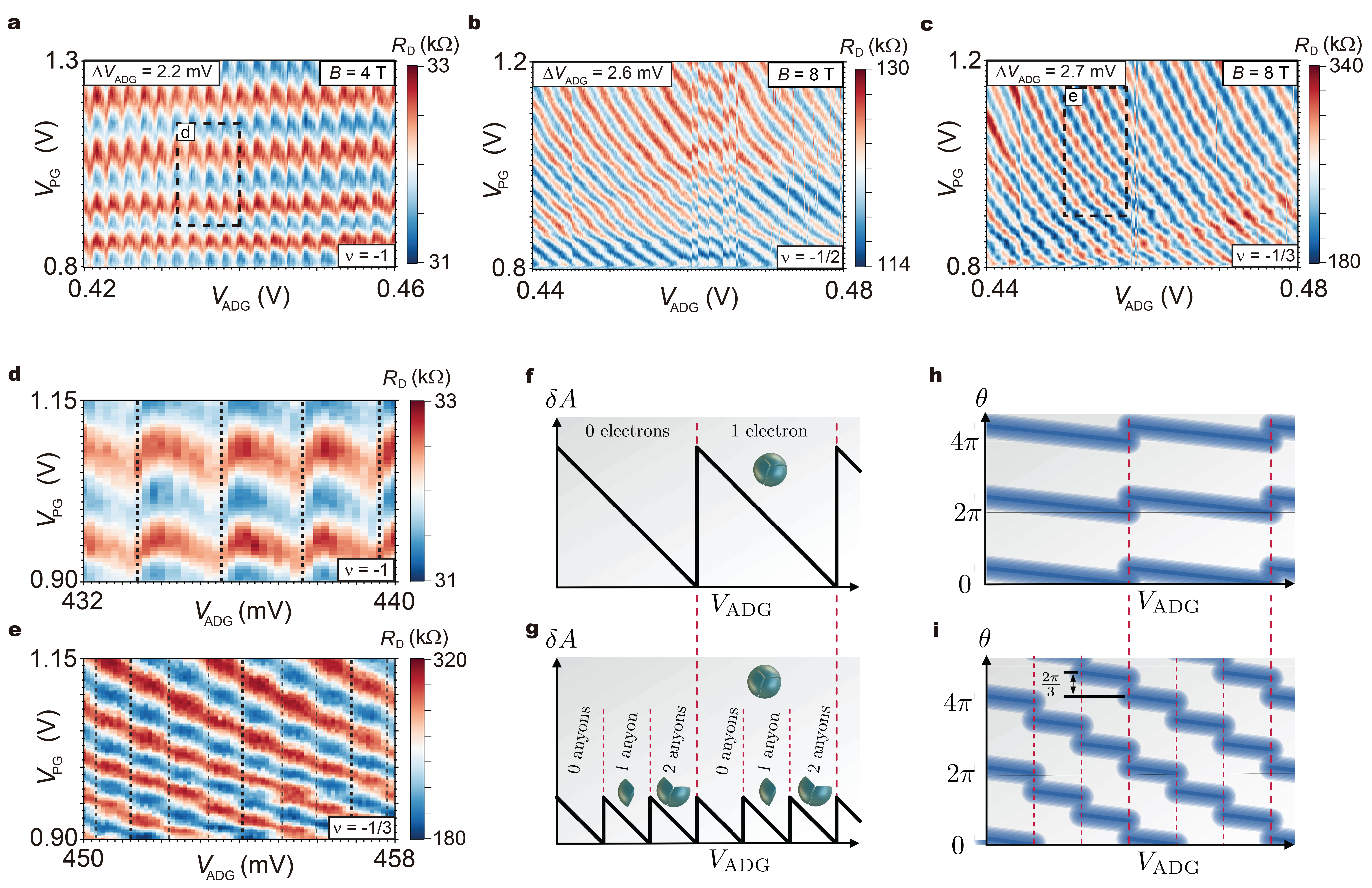}
  \caption{\textbf{Antidot-tuned interference in integer and fractional quantum Hall states.} (\textbf{a-c}) $R_\mathrm{D}$ displayed in the $V_\mathrm{ADG}$–$V_\mathrm{PG}$ plane at $\nu = -1$ under $B = 4~\mathrm{T}$ (a), $\nu = -\frac{1}{2}$ under $B = 8~\mathrm{T}$ (b), and $\nu = -\frac{1}{3}$ under $B = 8~\mathrm{T}$ (c). The indicated periods $\Delta V_\mathrm{ADG} $ were obtained from a 1D-FFT analysis; see SI4.  (\textbf{d, e}). Magnified views of $R_\mathrm{D}$ data from the dashed regions in (a) and (c), respectively. Vertical lines mark $V_\mathrm{ADG}$ values where phase discontinuities occur. (\textbf{f, g}) Schematic depiction of the interference area $\delta A$ changing continuously with $V_\mathrm{ADG}$ until a quantized electron or quasiparticle enters the interferometer. (\textbf{h, i}) The interference phase $\theta$ of Eq.~\eqref{eqn.theta} for the scenario illustrated in (f, g), agreeing qualitatively with the data in (d, e).}
  \label{fig2}
\end{figure}

To understand the effect of the antidot, we compare its influence for integers, odd-denominator, and even-denominator fractions. In Fig.~\ref{fig2}a-c, we show the interference patterns at $\nu=-1$, $-\frac{1}{2}$ and $-\frac{1}{3}$ as a function of $V_\mathrm{ADG}$ and $V_\mathrm{PG}$ at $\nu_\mathrm{AD}\approx 0$. In the integer case, shown in Fig.~\ref{fig2}a and magnified in Fig.~\ref{fig2}d, we observe a sawtooth shape of the constant-phase lines with a period of $\Delta V_\mathrm{ADG}\approx 2.2~\mathrm{mV}$ on top of a horizontal baseline. Using the capacitance $C_\mathrm{TG}$ = 0.647 mF/$\mathrm{m}^2$ obtained from the St$\check{\text{r}}$eda formula and the lithographic antidot area, this period corresponds to a charge difference of $\Delta Q \approx 0.89 e$ (see SI2). For a measurement made in a fixed magnetic field, a phase difference implies a change $\delta A$ in the interfering area. Such a change is expected for an incompressible quantum Hall liquid due to the Coulomb interactions with the ADG.\cite{halperin2011theory} At a critical point it becomes energetically favorable to change the liquid's total charge by adding or removing an electron, at which point the area resets, as illustrated in Fig.~\ref{fig2}f \cite{Ilani2004a, Ilani2004b}. This periodic change in area produces a sawtooth pattern in the interference phase; see Fig.~\ref{fig2}h. The horizontal baseline of the interference pattern in Fig.~\ref{fig2}a indicates an area change such that $\delta A B/\Phi_0 < \pi$.

\begin{figure}[H]
  \centering
  \includegraphics[width=.70\textwidth]{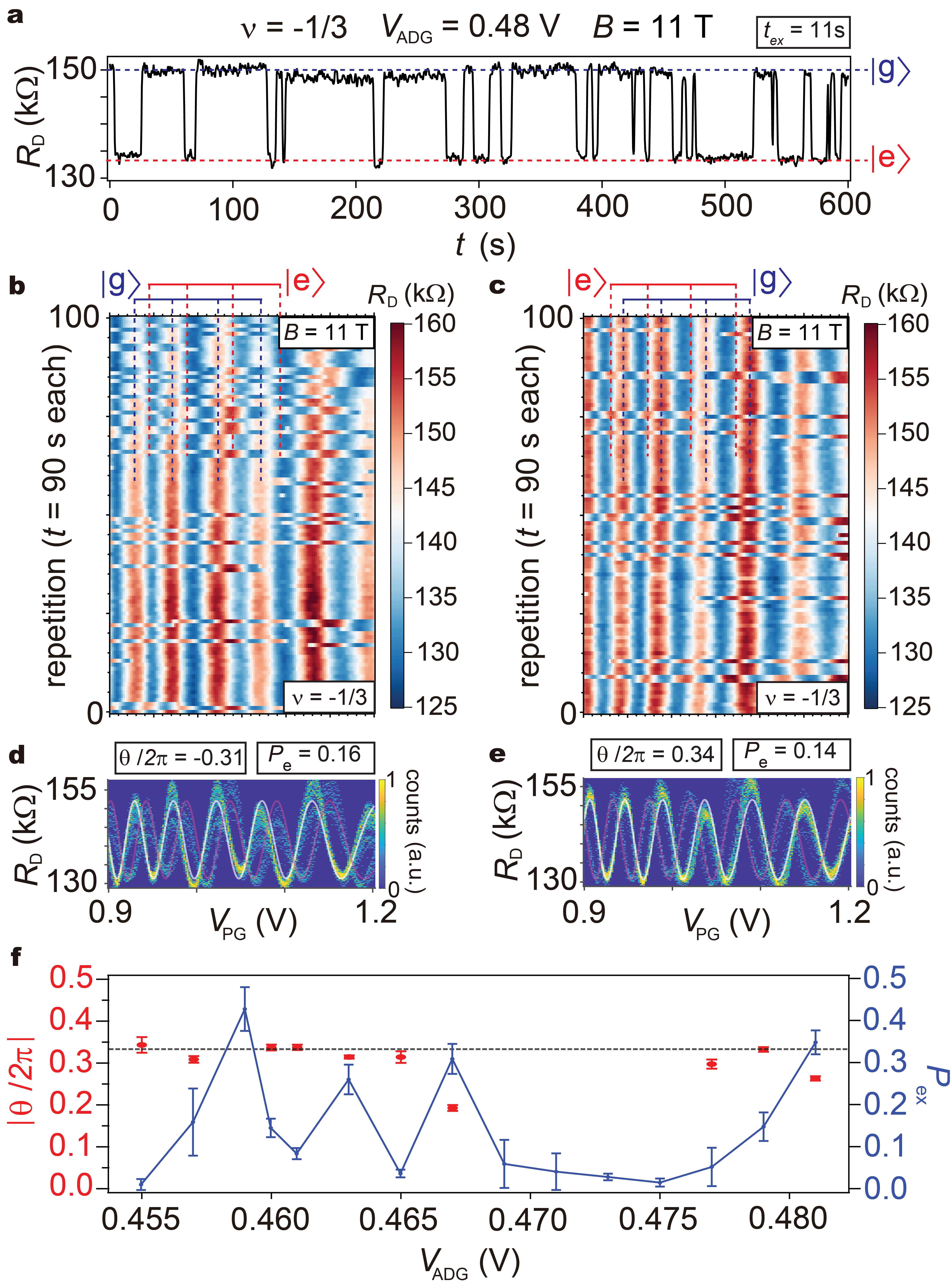}
  \caption{\textbf{Tunable anyon dynamics at $\nu=-\frac{1}{3}$}. (\textbf{a}) Time evolution of $R_\mathrm{D}$ at fixed magnetic field and gate voltages over 10 minutes. The excitation timescale $t_\mathrm{ex} \approx 11~\mathrm{s}$ was extracted from the autocorrelations of a 4000 seconds scan.  (\textbf{b}, \textbf{c}) $R_\mathrm{D}$ as a function of $V_\mathrm{PG}$, scanned 100 times at a rate of one sweep every 90 seconds, for $V_\mathrm{ADG}=0.457~\mathrm{V}$ and $0.46~\mathrm{V}$, both realizing $\nu_\mathrm{AD}=0$. (\textbf{d}, \textbf{e}) Histograms of $R_\mathrm{D}$ as a function of $V_\mathrm{PG}$, constructed from the data in (b, c).  (\textbf{f}) The excited state probability $P_\mathrm{ex}$ and the braiding phase $\theta_\mathrm{braid}$ as a function of $V_\mathrm{ADG}$, extracted from the histograms (see SI7 and SI9).}
  
  \label{fig3}
\end{figure}

\begin{figure}[H]
  \centering
  \includegraphics[width=.70\textwidth]{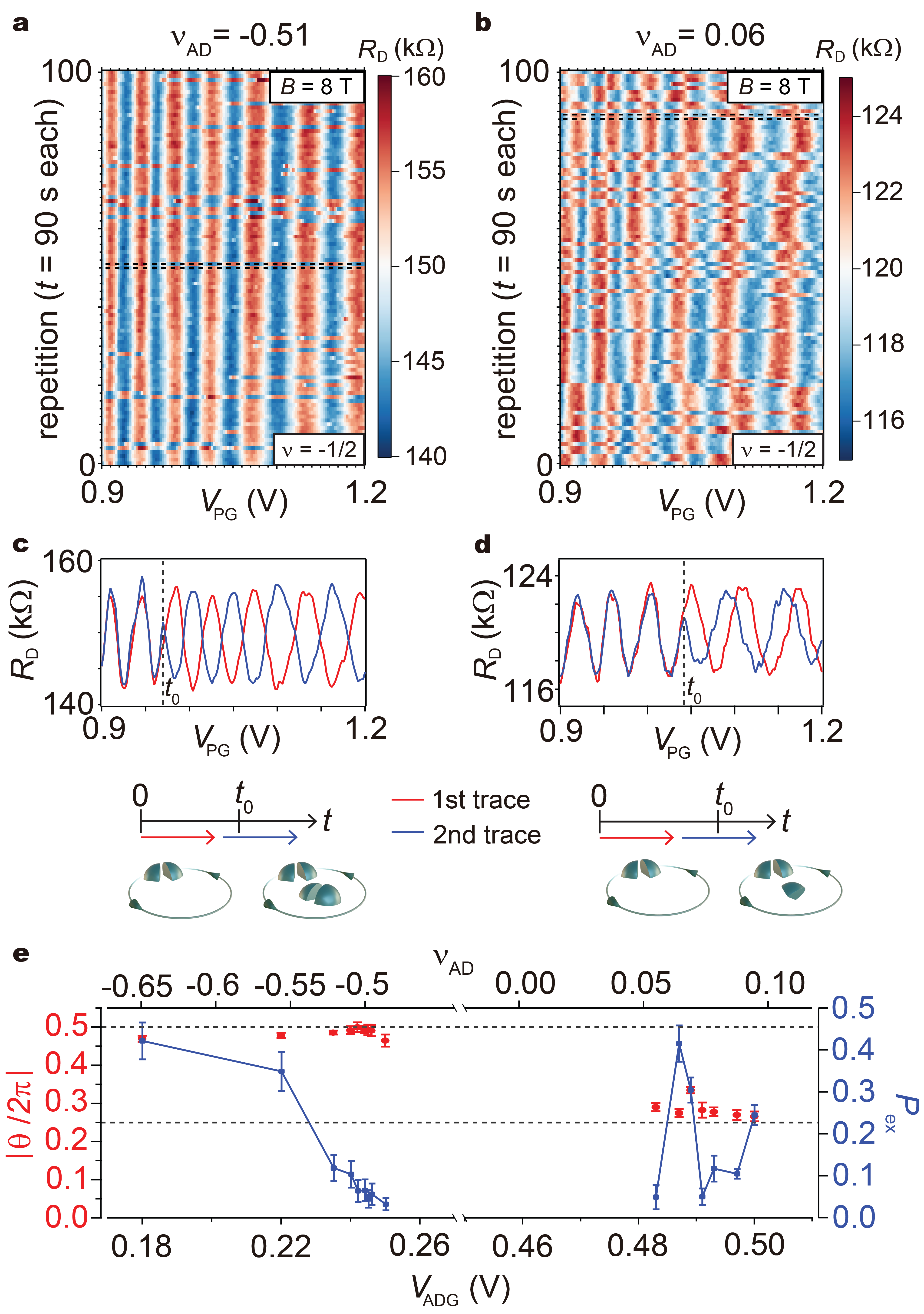}
  \caption{\textbf{Selective anyon dynamics at $\nu=-\frac{1}{2}$}.  (\textbf{a}, \textbf{b}) $R_\mathrm{D}$ as a function of $V_\mathrm{PG}$, scanned 100 times at a rate of one sweep every 90 seconds, for $V_\mathrm{ADG}=0.235~\mathrm{V}$, realizing $\nu_\mathrm{AD}=-\frac{1}{2}$,  and $0.487~\mathrm{V}$, realizing $\nu_\mathrm{AD}=0$. (\textbf{c}, \textbf{d}) Line scans of the consecutive $V_\mathrm{PG}$ sweeps indicated by dashed lines in (a, b). (\textbf{e}) The excited state probability $P_\mathrm{ex}$ and the braiding phase $\theta_\mathrm{braid}$ as a function of $V_\mathrm{ADG}$, extracted from the histograms (see SI8 and SI9).}
  \label{fig4}
\end{figure}

A different pattern appears at the two fractional fillings $\nu=-\frac{1}{2}$ and $-\frac{1}{3}$, which show tilted constant-phase baselines; see Figs.~\ref{fig2}b and c.  The ADG periodicities $\Delta V_\mathrm{ADG}\approx 2.6~\mathrm{mV}$ and $\Delta V_\mathrm{ADG}\approx 2.7~\mathrm{mV}$ are slightly larger than those in the integer case, but are consistent with the addition or removal of one electron. On top of the tilted baseline, two features occur. The first feature consists of abrupt, aperiodic phase jumps, which are the main focus of Figs.~\ref{fig3} and \ref{fig4}. The second are deviations from the straight baseline in the form of a staircase, most clearly visible at $\nu=-\frac{1}{3}$. A magnified view of this pattern reveals three steps within each $\Delta V_\mathrm{ADG}$ period, corresponding to $\Delta Q \approx e/3$; see Fig.~\ref{fig2}e. Similar to the integer case, the incompressible area changes continuously until a quantized charge is added. In contrast to the integer case, this charge is fractional at $\nu=-\frac{1}{3}$ and carried by an anyon with fractional braiding statistics; see Fig.~\ref{fig2}g. As a result, a single step in the staircase does not reset the interference phase, as it did for integers, but increments it by $\theta_\mathrm{braid}$, as illustrated in Fig.~\ref{fig2}i. Consistent with this picture, the phase differences between consecutive steps are $\Delta \theta / 2\pi = 0.33 \pm 0.02$; (see SI4).

At $\nu=-\frac{1}{2}$, the staircase pattern is almost completely smeared out. Similarly, the periodic steps at $\nu=-\frac{1}{3}$ are less prominent at different gate voltages (see SI4). In contrast, the aperiodic jumps are sharply defined and, in the case of $\nu=-\frac{1}{2}$, have not been reported before. The sharpness of the jumps directly reflects the lifetime of localized anyons inside the interference loop. Different anyon occupations are distinguishable and thus contribute incoherently (additively) to the resistance. Consequently, the staircase gets smeared out into an average slope if the number of localized anyons fluctuates substantially during the time resolution of the measurement.\cite{kim2026aharonov} This reasoning indicates slow anyon dynamics around the sharp jumps, as reported for $\nu=-\frac{1}{3}$ in,\cite{werkmeister2025anyon, samuelson2024anyonic} and prompts us to focus on the time dependence in the remainder of this work.

\noindent\textbf{Tunable anyon dynamics at $\nu=-\frac{1}{3}$}

Figure~\ref{fig3}a shows the time evolution of $R_\mathrm{D}$ at $\nu=-\frac{1}{3}$ and $\nu_\mathrm{AD}=0$ for constant gate voltages and a magnetic field of $B=11~\mathrm{T}$. $R_\mathrm{D}(t)$ was recorded for $4000~\mathrm{s}$ with a time resolution of $0.4~\mathrm{s}$ at a plunger gate set to a maximum of the time-averaged interference pattern. The data resemble the readout of a two-level system that briefly occupies an excited state before returning to the ground state via spontaneous emission. The two states correspond to different localized anyon occupations of $N$ and $N+1$, while the interfering anyon serves as the probe. From these jumps, we extract a lifetime of $t_\mathrm{ex}\approx 11~\mathrm{s}$ for the excited state.

Figures~\ref{fig3}b-c show $R_\mathrm{D}$ as a function of $V_\mathrm{PG}$ for two different $V_\mathrm{ADG}$ values for which $\nu_\mathrm{AD}=0$; no phase jumps were observed at antidot fillings $-\frac{1}{3} \leq \nu_\mathrm{AD}<0$ (see SI5). The plunger gate was swept for 100 time periods of $80~\mathrm{s}$ with a $10~\mathrm{s}$ return time. In this representation, the magnitude and sign of each phase jump are clearly visible. In particular, Fig.~\ref{fig3}b shows phase jumps of $\theta_\mathrm{braid}=-0.93\times \frac{2 \pi}{3} $, while Fig.~\ref{fig3}c shows  $\theta_\mathrm{braid}=+1.02\times \frac{2 \pi }{3}$. These values are consistent with an excited state with $\Delta N=1$ in the former case and $\Delta N=-1$ in the latter. We determine the probability $P_\mathrm{ex}$ to be in the excited state from the histograms in Fig.~\ref{fig3}d and e, finding $P_\mathrm{ex}=0.16$ and $P_\mathrm{ex}=0.14$, respectively.

Figure~\ref{fig3}f shows $P_\mathrm{ex}$ as a function of $V_\mathrm{ADG}$, extracted from the histograms (see SI7 and SI9). It exhibits three maxima with peak values of $30-50\%$, separated by $\Delta V_\mathrm{ADG} \approx5~\mathrm{mV}$. This behavior points to a near-degeneracy of the states containing $N$ and $N+1$ anyons at those $V_\mathrm{ADG}$ values. Close to these degeneracies, the sign of $\theta_\mathrm{braid}$, which is defined with respect to the ground state anyon occupation, becomes ambiguous. We, therefore, show $|\theta_\mathrm{braid}|$, finding values close to the expected value of $\frac{2\pi}{3}$. The phase jumps with different signs shown in Figs.~\ref{fig3}b,c occur on opposite sides of the first peak, but this pattern does not occur consistently for the other peaks (see SI5).

\noindent\textbf{Selective anyon dynamics at $\nu=-\frac{1}{2}$}

Our even-denominator interference measurements focus on $\nu=-\frac{1}{2}$, whose topological order, as suggested by nearby daughter states,\cite{huang2022valley, assouline2024energy, Hu2025, Kumar2025,levin2009collective,yutushui2024paired,zheltonozhskii2024identifying} is the Moore-Read Pfaffian.\cite{moore1991nonabelions} Its edge theory with $\nu=0$ is described by the anti-Pfaffian.\cite{lee2007particle,levin2007particle} Figures~\ref{fig4}a,b show $R_\mathrm{D}$ at $B=8~\mathrm{T}$, with $V_\mathrm{PG}$ swept as in Figs.~\ref{fig3}b,c.  The $V_\mathrm{ADG}$ value in Fig.~\ref{fig4}a realizes $\nu_\mathrm{AD}= \nu$, for which we observe distinct phase jumps, unlike the $\nu=-\frac{1}{3}$ case (see SI6). Together, these sweeps leads to the histogram in Fig.~\ref{fig1}f, from which we extracted $P_\mathrm{ex}=0.12$ and $\theta_\mathrm{braid} \approx 0.98 \pi$. Examining individual sweeps reveals phase jumps occurring over time. Figure~\ref{fig4}c shows line cuts around a particular phase jump where, after a time $t_0$, the entire interference pattern shifts by $\theta_\mathrm{braid} \approx \pi$. This observation indicates that one additional $e/2$ anyon has entered the interference loop at the time $t_0$. 

Figure~\ref{fig4}b presents data for $\nu_\mathrm{AD}=0$, where the phase jumps are notably smaller. The histogram in Fig.~\ref{fig1}g was constructed from this measurement and yielded $P_\mathrm{ex}=0.42$ and $\theta_\mathrm{braid} \approx 0.54 \pi$. Analysis of individual sweeps in Fig.~\ref{fig4}d shows that the interference pattern shifts rigidly, as in the previous case, but this time by the phase $\theta_\mathrm{braid} \approx \frac{\pi}{2}$. This finding indicates that one additional $e/4$ anyon has entered the interference loop and braided with the interfering $e/2$ anyons.

Figure~\ref{fig4}e shows $P_\mathrm{ex}$ and $|\theta_\mathrm{braid}|$ as a function of $V_\mathrm{ADG}$, extracted from the histograms (see SI8 and SI9). For $\nu_\mathrm{AD}= -\frac{1}{2}$, the phase jumps are very close to $\pi$ and independent of $P_\mathrm{ex}$, which ranges from $0$ to $50\%$. For $\nu_\mathrm{AD}=0$ the majority of phase jumps are close to $\frac{\pi}{2}$. Similar to the case of $\nu=-\frac{1}{3}$, excited states that are shifted in either direction, $\theta_\mathrm{braid} \approx \pm \frac{\pi}{2}$, with respect to the ground state are observed for different $V_\mathrm{ADG}$, but without  any clear correlation to $P_\mathrm{ex}$; see Fig. S11.

\color{black}

\newpage
\noindent\textbf{Conclusions}

The observation of non-Abelian braiding in interferometry experiments requires control over both the interfering and the localized anyons.\cite{stern2006proposed, bonderson2006detecting} Our work addresses one of these two challenges by establishing selectivity between localized $e/2$ and $e/4$ anyons in an even-denominator FQH state. Quasiparticles of either kind are trapped by tuning the gate voltage of an antidot inside the interferometer. We detect their presence through the observed braiding phases of $\theta_\mathrm{braid}=\pi$ or $\frac{\pi}{2}$ with the interfering $e/2$ anyons. Our time-dependent measurements resolve individual phase jumps due to either anyon type. Similar measurements at $\nu=-\frac{1}{3}$ confirm the expected statistics of $e/3$ anyons. Fluctuations in the number of localized anyons occur on long timescales at both fillings, over seconds to minutes, which could enable complex control and readout operations during their lifetimes.

Unlike the case of GaAs,\cite{ghosh2025coherent}, we did not observe any influence of the antidot on the charge of the interfering quasiparticles, a second requirement for observing non-Abelian braiding. Instead, the antidot filling strongly affects the observability of phase jumps, best seen at $\nu_\mathrm{ADG}=0$, at which there is a well-defined edge state between the antidot area and the bulk of the interferometer. Our observation of $\frac{\pi}{2}$ phase jumps occurring exclusively in this configuration suggests that a sharper potential profile at the QPCs could promote tunneling of $e/4$ particles there. Establishing control over the interfering anyon type, changing it from the currently realized $e/2$ quasiparticles to $e/4$, could complete the path toward the observations of non-Abelian braiding.

\pagebreak
\newpage
\section*{References}

%
\pagebreak
\section*{Materials and Methods}
\subsection{Stack Preparation $\colon$}
Van der Waals (vdW) heterostructures consisting of bilayer graphene encapsulated between hexagonal boron nitride (hBN) and graphite layers were used for device fabrication. Bulk graphite and hBN crystals were mechanically exfoliated onto SiO$_2$/Si substrates (10 mm × 10 mm) using adhesive tape. Suitable bilayer graphene, graphite and hBN flakes were identified by optical microscopy.
The heterostructure stack was assembled using a polycarbonate (PC) stamp mounted on a polydimethylsiloxane (PDMS) layer mounted on a glass slide. The PC film was heated at 170–180 °C for 2 h to improve adhesion to PDMS. The transfer stage was maintained at 130–131 °C during sequential pickup of the vdW layers in the following order: top graphite, top hBN, bilayer graphene, bottom hBN, and bottom graphite. The hBN thicknesses were 43 nm (top) and 30 nm (bottom).
The completed stack was transferred onto a clean SiO$_2$/Si substrate and heated at 180 °C for 15 min to melt the PC film. The PC was subsequently dissolved in chloroform for 3–4 h, followed by rinsing in isopropyl alcohol (IPA) and deionized water. The sample was annealed in ultrahigh vacuum (~10$^{-9}$ torr) at 400 °C for 4 h to remove residual contaminants and trapped bubbles.
Finally, contact-mode atomic force microscopy (AFM) ironing was performed with a 100 nN applied force to reduce strain fluctuations and flatten the stack on the atomic scale, thereby enhancing the energy gap of fractional quantum Hall states.

\subsection{Device fabrication $\colon$}
The bilayer-graphene-based Fabry--P\'erot interferometers (FPIs) were fabricated on five-layer vdW heterostructures assembled on highly p-doped Si substrates with a 280\,nm SiO$_2$ layer using standard nanofabrication techniques. Device geometry was defined by reactive ion etching (RIE), with polymethyl methacrylate (PMMA) as the etch mask. An O$_2$/CHF$_3$ mixture (1:10 volume ratio) was used for hBN etching, while O$_2$ plasma was used for graphite etching. After defining the geometry, the sample was annealed in ultrahigh vacuum ($\sim 10^{-9}$\,Torr) at 350$^\circ$C for 3\,h to remove resist residues.
A trench approximately 40\,nm wide was etched into the top graphite using mild O$_2$ plasma to minimize damage to the underlying hBN, thereby dividing the top graphite into seven independent sections. Each piece of the top graphite was connected to the voltage source via air bridges, enabling independent tuning of the electrostatic potential of each section. Bridge fabrication employed PMMA/MMA/PMMA trilayer resists, followed by a 20\,s mild O$_2$ plasma etch and evaporation of Cr (5\,nm)/Au (320\,nm).
In the final fabrication step, highly transparent edge contacts to the bilayer graphene were formed by first etching the top hBN layer using O$_2$/CHF$_3$ plasma, followed by angled evaporation of Cr (2\,nm)/Pd (20\,nm)/Au (60\,nm). This approach minimizes thermal exposure of the contacts during fabrication.

\subsection{Measurements$\colon$}
Measurements were performed in a dilution refrigerator with extensive filtering at a base temperature of 10 mK using a standard low-frequency lock-in technique. An SRS 865A lock-in amplifier was used to generate an alternating voltage at 11.7 Hz and measure the voltage difference between two contacts with a 300 ms time constant. A 100 M$\Omega$ load resistor was included in series with the lock-in amplifier, resulting in an alternating current ranging from 50 nA to 0.5 nA.
A QDAC, an ultralow-noise 24-channel digital-to-analog converter (QDevil-QM), was used to tune the voltages applied to all graphite gates. Additionally, a Keithley 2400 voltage source was used to apply a voltage to the highly p-doped Si substrate, doping the contact region and improving the contact resistance.

\section*{Data and materials availability:}
The data supporting the plots in this paper and other findings of this study are available from the corresponding author upon request.

 \section*{Acknowledgements}
It is a pleasure to thank Moty Heiblum for illuminating discussions. \textbf{Funding:} J.K. acknowledges support from the Dean of the Faculty and the Clore Foundation. Y.R. acknowledges the support from the Quantum Science and Technology Program 2021 and the Schwartz Reisman Collaborative Science Program; the Shimon and Golde Picker–Weizmann Annual Grant; research grants from the Goldfield Family Charitable Trust, the Estate of Hermine Miller, and the Sheba Foundation, Dweck Philanthropies, Inc.; the European Research Council Starting Investigator Grant No. 101163917; the Minerva Foundation with funding from the Federal German Ministry for Education and Research; the Israel Science Foundation under grants No. 380/25 and 425/25; and the DFG (CRC/Transregio 183). D.F.M acknowledges support from the ISF (grant 3281/25); the Minerva Foundation with funding from the Federal German Ministry for Education and Research; and the DFG (CRC/Transregio 183). A.S. acknowledges support from the ISF, ISF Quantum Science and Technology (grant 2074/19); and the DFG (CRC/Transregio 183).

\section*{Author contributions}
J.K. prepared the stacks. K.W. and T.T. provided the hBN crystals. J.K. and A.I. improved the device quality. J.K. fabricated the device. J.K. and R.K. developed the measurement circuit and dilution refrigerator. J.K. performed the measurements. J.K., A. Shaer, A. Stern, D.F.M., and Y.R. analyzed the data. A. Shaer, A. Stern, and D.F.M. developed the theoretical framework. J.K., A. Shaer, A. Stern, D.F.M., and Y.R. wrote the manuscript with input from all authors. Y.R. supervised the project.

\section*{Competing interests}
The authors declare no competing interests.
\newpage


\section*{Supplementary material}

# Supplementary Information for

# Selective Braiding of different Anyons in the Even-Denominator Fractional Quantum Hall Effect

# Contents

## SI1  QPC characterization

In this section, we characterize the left and right quantum point contacts (LQPC and RQPC) for all studied filling factors. In our device, the QPC saddle-point potentials are screened by the top graphite layer. As a result, the two air bridges (LBG and RBG) cannot effectively control the QPC transmission. Instead, each QPC is formed by tuning the filling factor beneath the LSG or RSG. This tuning determines the QPC transmission, $t_{\text{QPC}} = G_{\text{D}} h/e^2$, where the diagonal conductance is defined as $G_{\text{D}} = 1/R_{\text{D}}$. We define the relative transmission at the left (right) QPC as $t_{L(R)} = G_{\text{D}}/G_{xy}$ where $G_{xy} = \nu e^2/h$ is the bulk Hall conductance. $G_{\text{D}}$ is measured while operating either the left or right QPC, with the other kept open.

Figure S1 shows the diagonal conductance $G_{\text{D}}$ for LQPC (red curve) and RQPC (blue curve), measured as a function of $V_{\text{LSG}}$ or $V_{\text{RSG}}$ at different filling factors, with a bias current of $I_{\text{SD}} = 50$ pA. When the edge is fully transmitted across the QPC, $G_{\text{D}}$ approaches $G_{xy}$, as indicated by plateau behavior at $\nu = -1$ and $\nu = -1/3$. However, for other fractional fillings ($\nu = -2/3$ and $\nu = -1/2$), $G_{\text{D}}$ is smaller than $G_{xy}$ even when the QPCs are fully opened. This reduction may arise because the QPCs are too narrow relative to the width of the fractional edge, an effect that may be particularly important for edges with a more complex structure. In our measurements, $t_{\text{L}}$ and $t_{\text{R}}$ are set to 0.6–0.8 for interference at integer filling factors, and 0.7–0.9 for interference at the fractional filling factors.

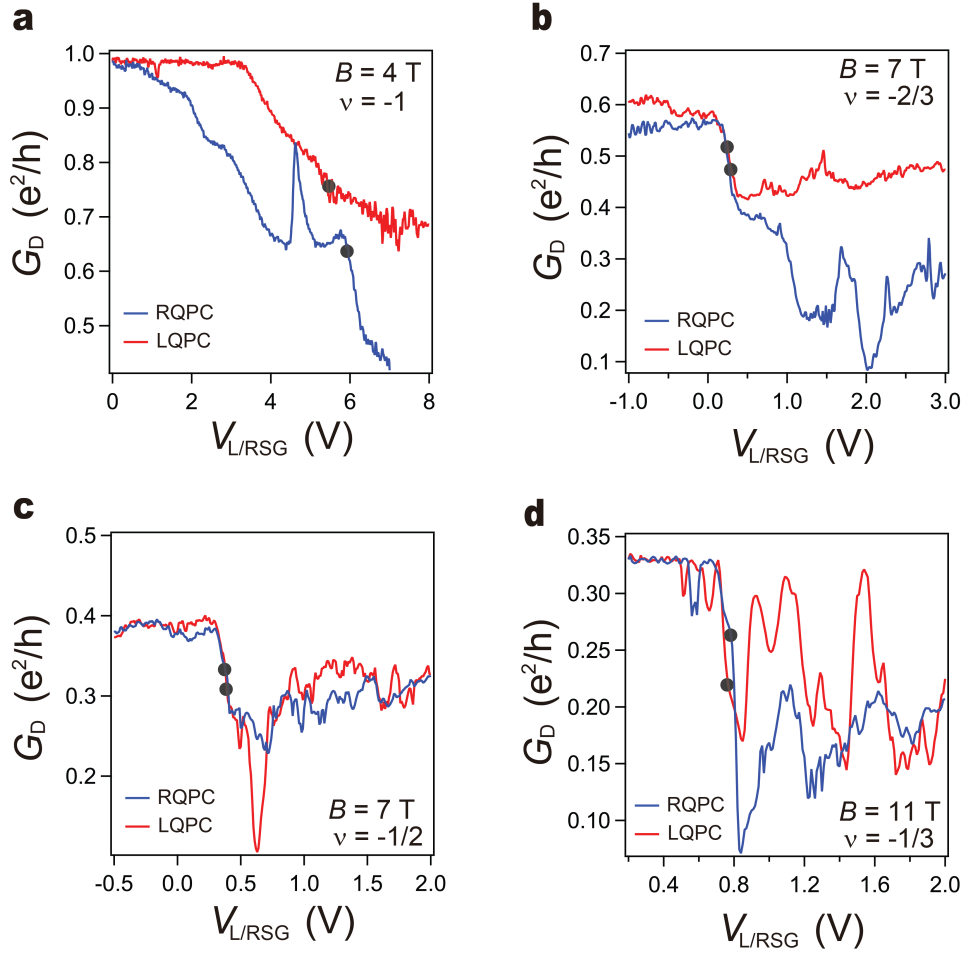

**Figure S1: QPC characterization for integer and fractional fillings. (a-d)** Diagonal conductance $G_\mathrm{D}$ for LQPC (red) and RQPC (blue), measured as a function of $V_\mathrm{LSG}$ or $V_\mathrm{RSG}$ at different fillings $\nu$ in perpendicular magnetic fields ranging from 4 to 11 T. **(a)** $\nu = -1$. **(b)** $\nu = -\frac{2}{3}$. **(c)** $\nu = -\frac{1}{2}$. **(d)** $\nu = -\frac{1}{3}$. Black dots on each curve indicate the typical operating points used for the interference measurements. A single ground contact is used for all $G_\mathrm{D}$ measurements.

## SI2 Extracting $\alpha_c$ from longitudinal resistance measurements.

In this study, we measure Aharonov–Bohm oscillations by monitoring $R_\text{D}$ under conditions where the filling factor is maintained while varying the magnetic field, following our previous approach.[1] We present the $R_\text{D}$ data in a two-dimensional plane, defined by trajectories of constant $\alpha \equiv dB/dV_\text{TG}$ and $V_\text{PG}$. To observe only the Aharonov-Bohm contribution, $V_\text{TG}$ and $V_\text{ADG}$ are varied together with $B$ to maintain a constant filling factor in the region enclosed by the interfering quasiparticles, satisfying $\alpha_c = \frac{\delta B}{\delta V_\text{TG(ADG)}}$. To determine the constant-filling trajectory $\alpha_c$ used in the measurements of Aharonov-Bohm oscillations, we analyze the longitudinal resistance $R_\text{xx}$ fan diagram measured in the $B$–$V_\text{TG}$ plane. Figure S2a shows $R_\text{xx}$ as a function of $V_\text{TG}$ and $B$ at a fixed $V_\text{BG} = -0.3V$, confirming $\nu = -2/3, -1/2$, and $-1/3$ fractional quantum Hall (FQH) states in bilayer graphene. Figure S2b shows $R_\text{xx}$ line cuts taken at $B$ = 6.5 T (blue curve) and $B$ = 7.5 T (red curve), taken from Fig. S2a. The shift in $V_\text{TG}$ is determined by calculating the difference between the positions of two vertical dashed lines drawn at the center of the $R_\text{xx}$ dip. This corresponds to 40 mV for $\nu = -2/3$, 29.8 mV for $\nu = -1/2$, and 20 mV for $\nu = -1/3$. Using these values, we calculate $\alpha_c = \frac{\delta B}{\delta V_\text{TG}}$. The FQH states at $\nu = -2/3, -1/2$, and $-1/3$ yield $\alpha_c$ values of $-25$ T/V, $-33.5$ T/V, and $-50$ T/V, respectively. In Fig. S2c, we plot the extracted $\alpha_c$ as a function of $1/\nu$, consistent with the Streda formula. From the Streda formula $C_\text{TG} = \frac{e\alpha_c\nu}{\Phi_0}$, the capacitance of the top gate (TG) per unit area is estimated as $C_\text{TG}$ = 0.647 mF/m$^2$, using the average of $\alpha_c\nu$ obtained from the linear fit in Fig. S2c (16.7 T/V).

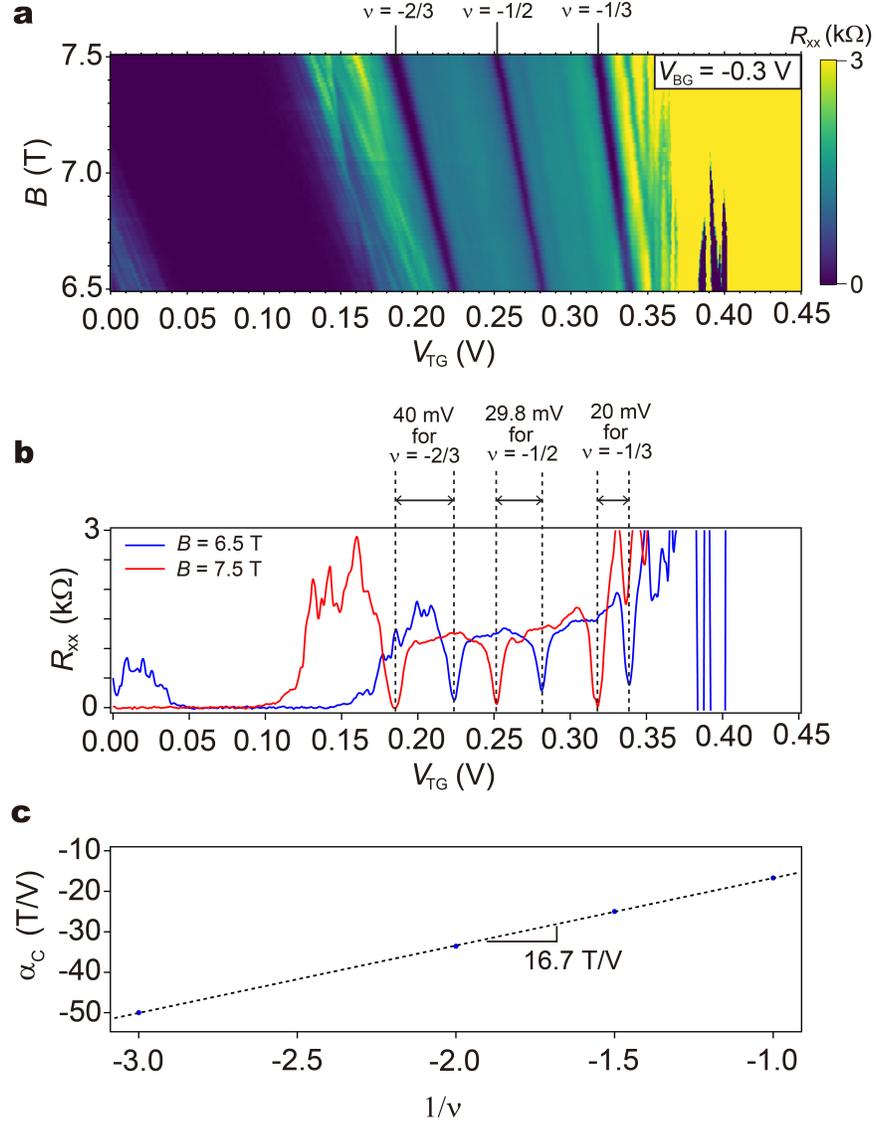

**Figure S2: Determination of $\alpha_C$ values.** **(a)** $R_{xx}$ as a function of $V_{TG}$ and $B$ at a fixed $V_{BG} = -0.3$ V. **(b)** Line cuts of $R_{xx}$ as a function of $V_{TG}$ at $B$ = 6.5 T (blue) and $B$ = 7.5 T (red). **(c)** Extracted $\alpha_c$ as a function of $1/\nu$. The black dashed line shows a linear fit to $\alpha_c(1/\nu)$.

4

## SI3  Aharonov–Bohm interference

Figures S3a-d show $\Delta R_\mathrm{D}$ measurements plotted in $B|_{\alpha_c}$-$V_\mathrm{PG}$ planes at filling factors of $\nu = -1, -2/3, -1/2$, and $-1/3$. For all fillings, clear AB-dominated interference patterns are observed. For hole doping, constant-phase lines in the AB regime are expected to exhibit positive slopes, unlike the negative slopes observed for electron doping. Figures S3e-h show the 2D FFTs of the corresponding data in Figs. S3a-d, respectively, plotted as a function of $\frac{\Phi_0}{\Delta B}$ and $\frac{1}{\Delta V_\mathrm{PG}}$. To determine the magnetic field periodicity $\Delta B$, we analyze one-dimensional cuts of the 2D-FFTs as a function of $\frac{\Phi_0}{\Delta B}$, taken at the peak value of $\Delta V_\mathrm{PG}$. Figures S3i-l show the FFT line cuts as a function of $\frac{\Phi_0}{\Delta B}$. By fitting a Gaussian curve to the data, we extract the mean value $\frac{\Phi_0}{\Delta B_\mathrm{fit}}$ and the variance $\sigma^2$ listed in the lower-right corners of Figs. S3i-l.

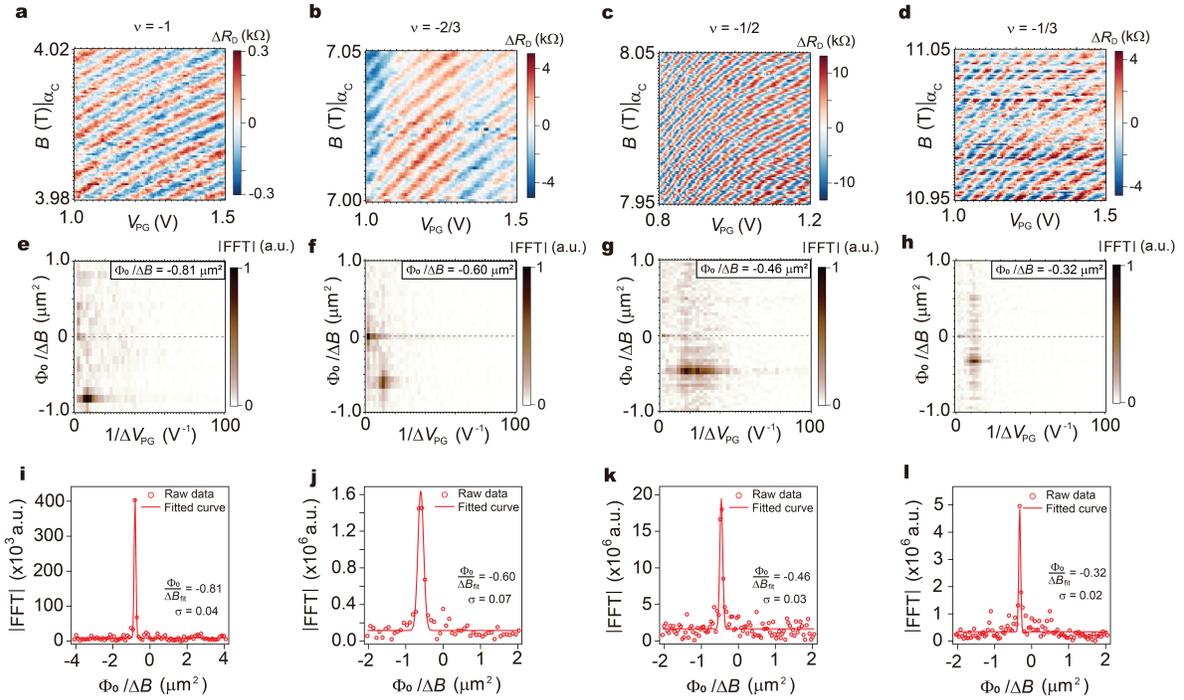

**Figure S3: Aharonov–Bohm interference.** (**a-d**) $\Delta R_\mathrm{D}$ for $\nu = -1, -2/3, -1/2$, and $-1/3$, measured in the $B|_{\alpha_c}$-$V_\mathrm{PG}$ plane. (**e-h**) 2D-FFT as a function of $\frac{\Phi_0}{\Delta B}$ and $\frac{1}{\Delta V_\mathrm{PG}}$, corresponding to Figs. S3a-d, respectively. (**i-l**) One-dimensional cuts of the 2D-FFT as a function of $\frac{\Phi_0}{\Delta B}$, taken at the peak value of $\Delta V_\mathrm{PG}$, obtained from Figs. S3e-h, respectively. Lower-right corners show the Gaussian fitting parameters.

## SI4 Phase extraction from one-dimensional Fourier analysis

To extract the phase of the $R_\mathrm{D}$ oscillations, we performed a one-dimensional fast Fourier transform (FFT) along the $V_\mathrm{PG}$ direction for each fixed $V_\mathrm{ADG}$.[2,3] The phase $\theta(V_\mathrm{ADG})$ is defined as the argument of the complex FFT coefficient at the dominant oscillation frequency,

$$\theta(V_\mathrm{ADG}) = \arg\left[\tilde{S}(f_0; V_\mathrm{ADG})\right],$$

where $f_0$ denotes the oscillation frequency obtained from the peak of the FFT magnitude spectrum. The resulting phase was unwrapped along $V_\mathrm{ADG}$ to remove $2\pi$ discontinuities. This FFT-based procedure provides a direct estimate of the phase evolution as a function of $V_\mathrm{ADG}$.

The results of this analysis are shown in Fig. S4. In Fig. S4a-c, the upper panels display $R_\mathrm{D}$ in the $V_\mathrm{ADG}$–$V_\mathrm{PG}$ plane for $\nu = -1, -1/2$, and $-1/3$, respectively. The lower panels show the phase $\theta/2\pi$ extracted from the FFT analysis described above, which we use to determine $\Delta V_\mathrm{ADG}$. In Fig. S4a, $\theta$ is a periodic function of $V_\mathrm{ADG}$, and the spacing between successive local maxima yields $\Delta V_\mathrm{ADG} \approx 2.2$ mV. In Fig. S4b, $\theta$ changes monotonically as a function of $V_\mathrm{ADG}$, allowing $\Delta V_\mathrm{ADG} \approx 2.6$ mV to be extracted directly, which corresponds to the change in $V_\mathrm{ADG}$ required for a $2\pi$ phase accumulation. In Fig. S4c, $\theta$ exhibits a nearly linear dependence on $V_\mathrm{ADG}$, with step-like phase changes indicated by the dashed lines. The dashed lines correspond to local minima of the derivative $d\theta/dV_\mathrm{ADG}$ obtained from the phase in Fig. S4c. As in Fig. S4b, $\Delta V_\mathrm{ADG} \approx 2.7$ mV is obtained.

To quantify the phase differences $\Delta\theta$ between consecutive steps in Fig. S4c, we analyzed the derivative of the phase trace with respect to $V_\mathrm{ADG}$. We identify local minima in $d\theta/dV_\mathrm{ADG}$ as characteristic points marking the centers of the phase-slip events. The positions of these minima define a set of $V_\mathrm{ADG}$ values, from which the spacing $\Delta V_\mathrm{ADG}$ between neighboring events was obtained. The phase difference $\Delta\theta$ between consecutive minima was then calculated from the phase values at these positions. From this distribution, we obtain $\Delta\theta/2\pi = 0.33 \pm 0.02$.

Figures S4e and S4f display an extended $V_\mathrm{ADG}$ range and a magnified view of the dashed region, respectively, highlighting that the periodic steps at $\nu = -\frac{1}{3}$ become less pronounced at different $V_\mathrm{ADG}$ regions.

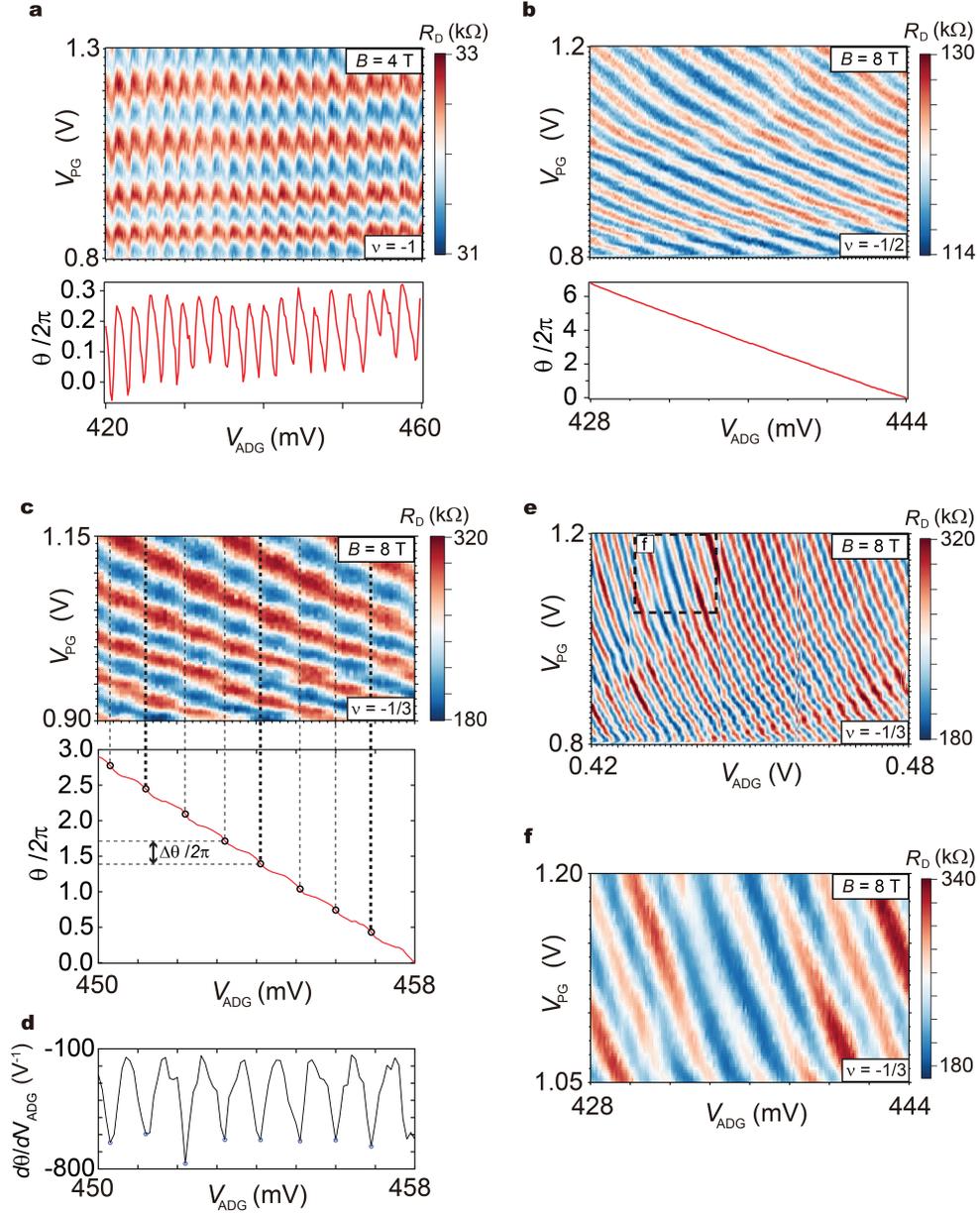

**Figure S4: Phase extraction from one-dimensional Fourier analysis.** (a-c) The upper panels show $R_D$ displayed in the $V_{ADG}$–$V_{PG}$ plane at $\nu = -1$ under $B = 4$ T (a), $\nu = -\frac{1}{2}$ under $B = 8$ T (b), and $\nu = -\frac{1}{3}$ under $B = 8$ T (c). The lower panel shows the phase $\theta/2\pi$ extracted from a one-dimensional FFT performed along the $V_{PG}$ direction for each fixed $V_{ADG}$. (d) Derivative $d\theta/dV_{ADG}$ obtained from the phase in (c). (e) $R_D$ displayed in the $V_{ADG}$–$V_{PG}$ plane at $\nu = -\frac{1}{3}$ under $B = 8$ T with a larger $V_{ADG}$ range. (f) Magnified view of the dashed region in (e).

## SI5 Repeated measurements of $R_\mathrm{D}$ at $\nu = -1/3$ under 11 T

Figures S5 and S6 show the full set of repeated measurements of $R_\mathrm{D}$ as a function of $V_\mathrm{PG}$ at $\nu = -1/3$ under a fixed magnetic field of 11 T, as discussed in the main text. We used an SRS 865A lock-in amplifier to generate an AC excitation current of 50 pA at 11.7 Hz and measure the diagonal voltage with a 300 ms time constant. Each data point requires approximately 400 ms, including the time to change $V_\mathrm{PG}$ and for data acquisition. Figure S5 shows the repeated measurements of $R_\mathrm{D}$ for $-\frac{1}{3} \leq \nu_\mathrm{AD} < 0$, where no phase jumps are observed. Straight lines of constant phase indicate that our measurement conditions remain stable for more than 2 hours. Some traces show a slight change in visibility, but the the phase remains stable over the measurement time. Figure S6 shows the repeated measurements of $R_\mathrm{D}$ for $\nu_\mathrm{AD} > 0$, where phase jumps are clearly observed. The phase jumps occur randomly relative to the straight lines of constant phase. We used these data to construct the histograms of $R_\mathrm{D}$ shown in the main text.

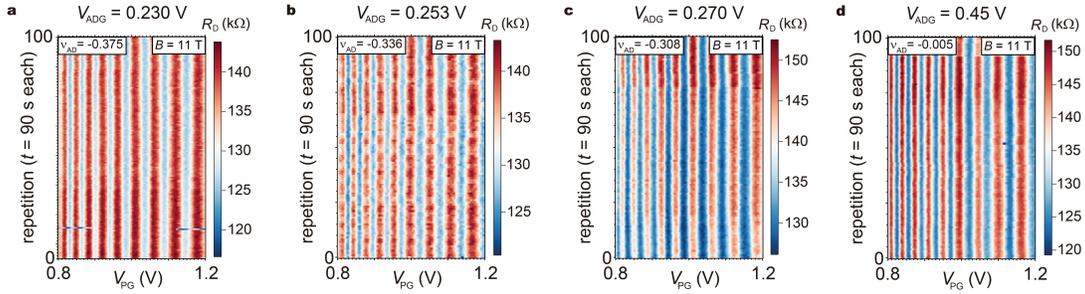

**Figure S5: Repeated measurements of $R_\mathrm{D}$ as a function of $V_\mathrm{PG}$ for $-\frac{1}{3} \leq \nu_\mathrm{AD} < 0$.** (a-d) $R_\mathrm{D}(V_\mathrm{PG})$ repeated 100 times with a sweep period of 90 s. (a) $\nu_\mathrm{AD} = -0.375$. (b) $\nu_\mathrm{AD} = -0.336$. (c) $\nu_\mathrm{AD} = -0.308$. (d) $\nu_\mathrm{AD} = -0.005$. $V_\mathrm{ADG}$ values corresponding to each $\nu_\mathrm{AD}$ are indicated above each panel.

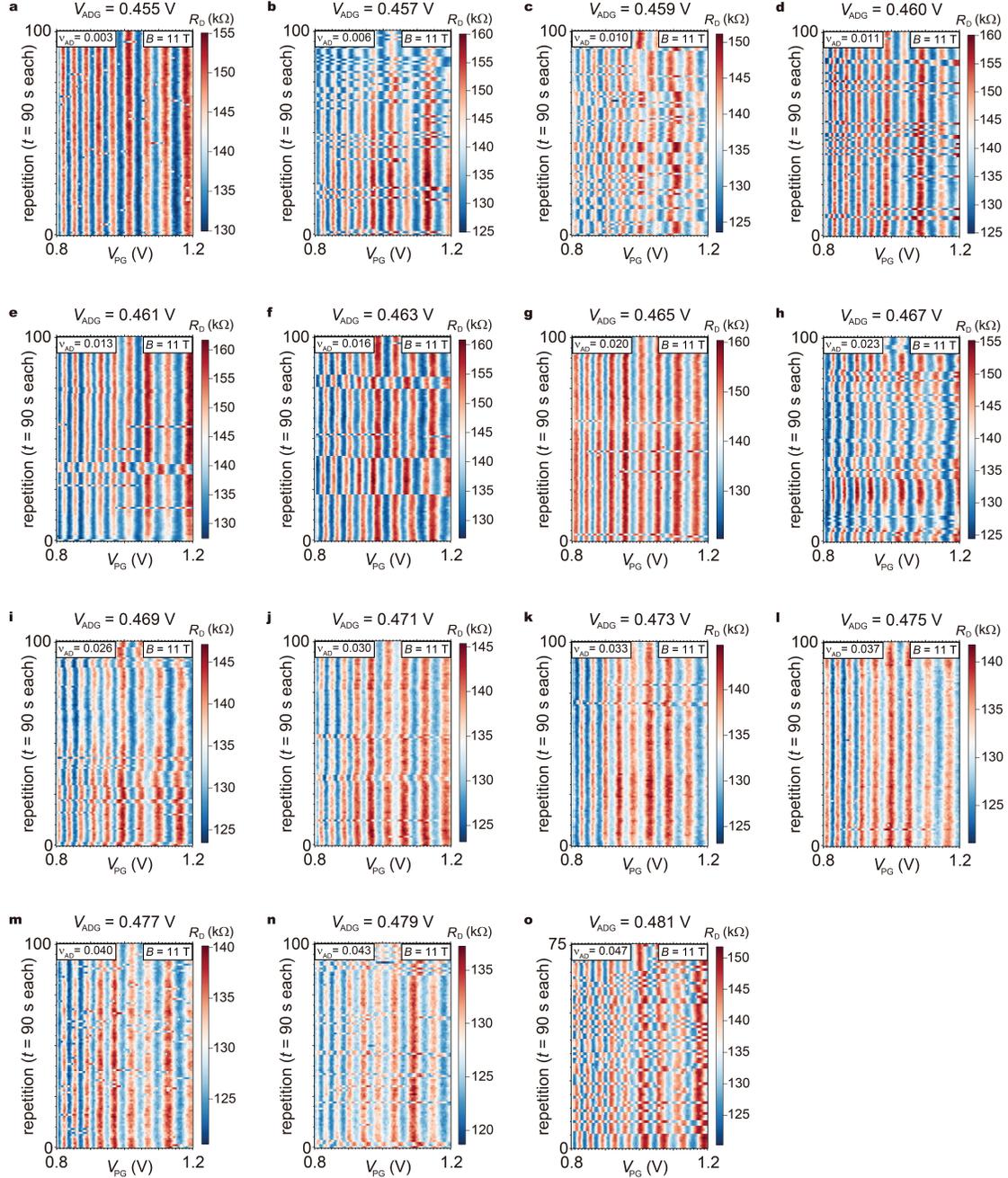

**Figure S6: Repeated measurements of $R_\mathrm{D}$ as a function of $V_\mathrm{PG}$ for $\nu_\mathrm{AD} > 0$. (a-o)** $R_\mathrm{D}$ ($V_\mathrm{PG}$) repeated 100 times with a sweep period of 90 s, except for panel (o), which was repeated 75 times. **(a)** $\nu_\mathrm{AD} = 0.003$. **(b)** $\nu_\mathrm{AD} = 0.006$. **(c)** $\nu_\mathrm{AD} = 0.010$. **(d)** $\nu_\mathrm{AD} = 0.011$. **(e)** $\nu_\mathrm{AD} = 0.013$. **(f)** $\nu_\mathrm{AD} = 0.016$. **(g)** $\nu_\mathrm{AD} = 0.020$. **(h)** $\nu_\mathrm{AD} = 0.023$. **(i)** $\nu_\mathrm{AD} = 0.026$. **(j)** $\nu_\mathrm{AD} = 0.030$. **(k)** $\nu_\mathrm{AD} = 0.033$. **(l)** $\nu_\mathrm{AD} = 0.037$. **(m)** $\nu_\mathrm{AD} = 0.040$. **(n)** $\nu_\mathrm{AD} = 0.043$. **(o)** $\nu_\mathrm{AD} = 0.047$. $V_\mathrm{ADG}$ values corresponding to each $\nu_\mathrm{AD}$ are indicated above each panel.

9

## SI6 Repeated measurements of $R_D$ at $\nu = -1/2$ under 8 T

Figures S7 and S8 show the full set of repeated measurements of $R_D$ as a function of $V_{PG}$ at $\nu = -1/2$ under a fixed magnetic field of 8 T, discussed in the main text. We used an SRS 865A lock-in amplifier to generate an AC excitation current of 50 pA at 11.7 Hz and read out the diagonal voltage with a 300 ms time constant. Each data point requires approximately 400 ms, including the time to change $V_{PG}$ and for data acquisition. Figure S7 shows the repeated measurements of $R_D$ for $\nu_{AD} < 0$. As $\nu_{AD}$ decreases below $\nu_{AD} = -1/2$, the number of phase jumps tends to increase. However, as $\nu_{AD}$ increases above $\nu_{AD} = -1/2$, the number of phase jumps decreases and is completely suppressed when $\nu_{AD}$ approaches $\nu_{AD} = 0$. The phase jumps start to appear again when $\nu_{AD} > 0$, as shown in Fig. S8. We used these data to construct the histograms of $R_D$ shown in the main text.

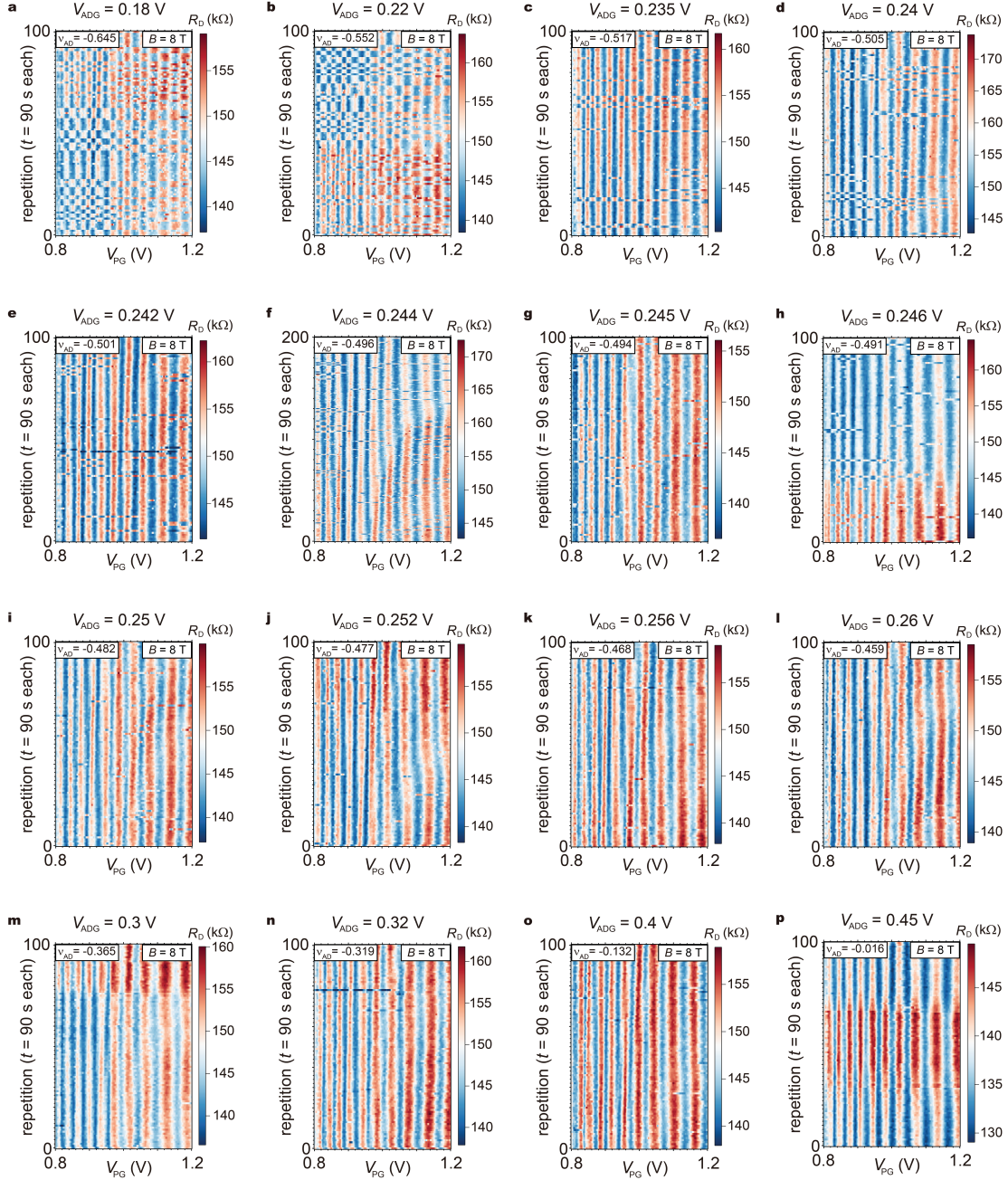

**Figure S7: Repeated measurements of $R_D$ as a function of $V_{PG}$ for $\nu_{AD} < 0$.** (a-p) $R_D$ ($V_{PG}$) repeated 100 times with a sweep period of 90 s. (a) $\nu_{AD} = -0.645$. (b) $\nu_{AD} = -0.552$. (c) $\nu_{AD} = -0.517$. (d) $\nu_{AD} = -0.505$. (e) $\nu_{AD} = -0.501$. (f) $\nu_{AD} = -0.496$. (g) $\nu_{AD} = -0.494$. (h) $\nu_{AD} = -0.491$. (i) $\nu_{AD} = -0.482$. (j) $\nu_{AD} = -0.477$. (k) $\nu_{AD} = -0.468$. (l) $\nu_{AD} = -0.459$. (m) $\nu_{AD} = -0.365$. (n) $\nu_{AD} = -0.319$. (o) $\nu_{AD} = -0.132$. (p) $\nu_{AD} = -0.016$. $V_{ADG}$ values corresponding to each $\nu_{AD}$ are indicated above each panel.

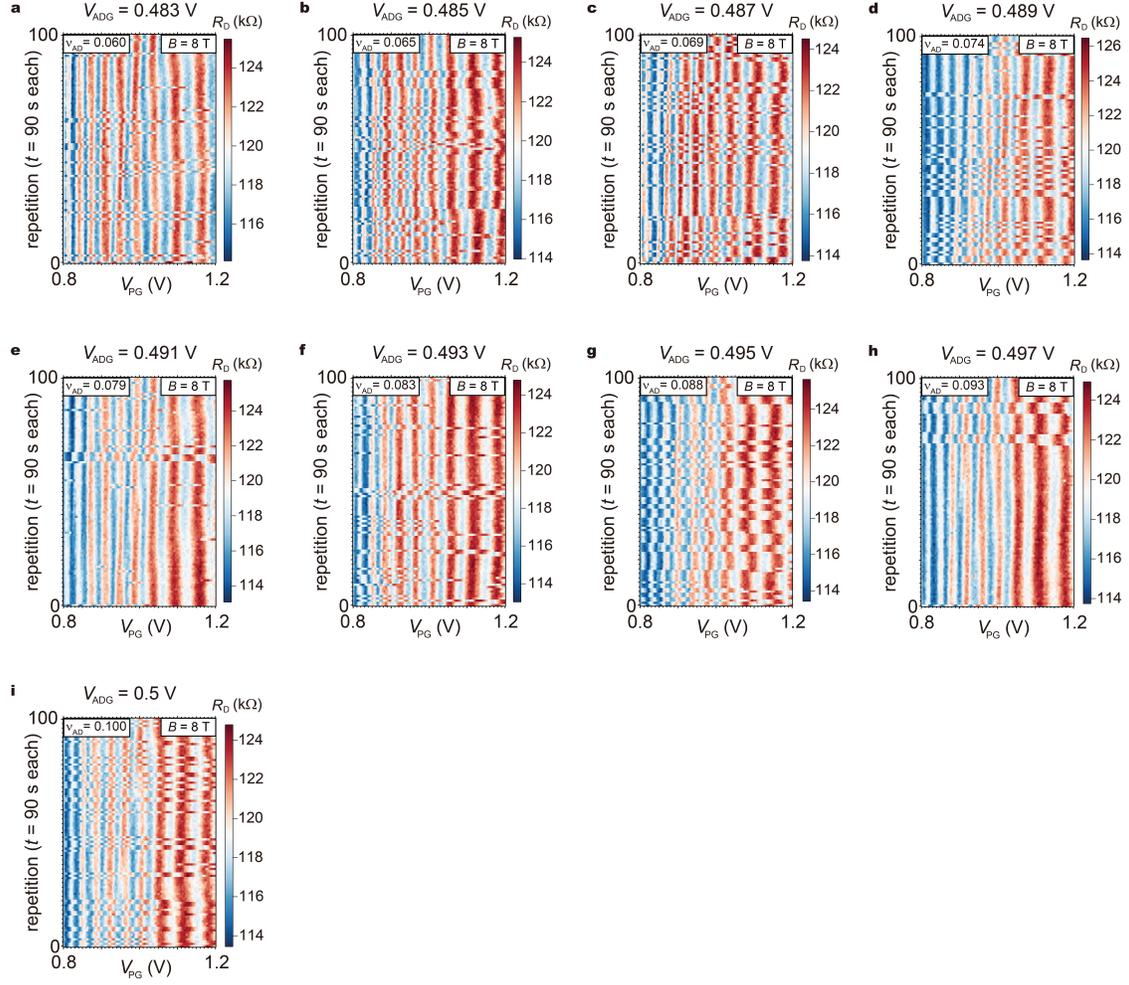

**Figure S8: Repeated measurements of $R_\mathrm{D}$ as a function of $V_\mathrm{PG}$ for $\nu_\mathrm{AD} > 0$.** (a-i) $R_\mathrm{D}(V_\mathrm{PG})$ repeated 100 times with a sweep period of 90 s. (a) $\nu_\mathrm{AD} = 0.060$. (b) $\nu_\mathrm{AD} = 0.065$. (c) $\nu_\mathrm{AD} = 0.069$. (d) $\nu_\mathrm{AD} = 0.074$. (e) $\nu_\mathrm{AD} = 0.079$. (f) $\nu_\mathrm{AD} = 0.083$. (g) $\nu_\mathrm{AD} = 0.088$. (h) $\nu_\mathrm{AD} = 0.093$. (i) $\nu_\mathrm{AD} = 0.100$. $V_\mathrm{ADG}$ values corresponding to each $\nu_\mathrm{AD}$ are indicated above each panel.

12

## SI7 Histograms of $R_D$ at $\nu = -1/3$ under 11 T

We construct two-dimensional histograms of $R_D$ at each $V_{PG}$ by binning repeated measurements, with the bin range defined by the 1st–99th percentiles of the data to suppress outliers [4]. We identify the primary ridge of the histogram by locating the maximum of a smoothed histogram slice. The ridge was fitted to a cosine function with a polynomial phase $p(V_{PG})$. A second ridge was identified from additional peaks in the slices that were well separated from the primary ridge. The two ridges were then fitted using a common amplitude and polynomial phase while allowing a relative phase offset. The fitted curves are shown in Fig. S9, with the white curve for the primary ridge and the red curve for the secondary ridge. The phase difference between the two fitted curves was determined from this fit and is shown in the upper-left corner of each panel. We estimate the uncertainties from the residual variance of this fit.

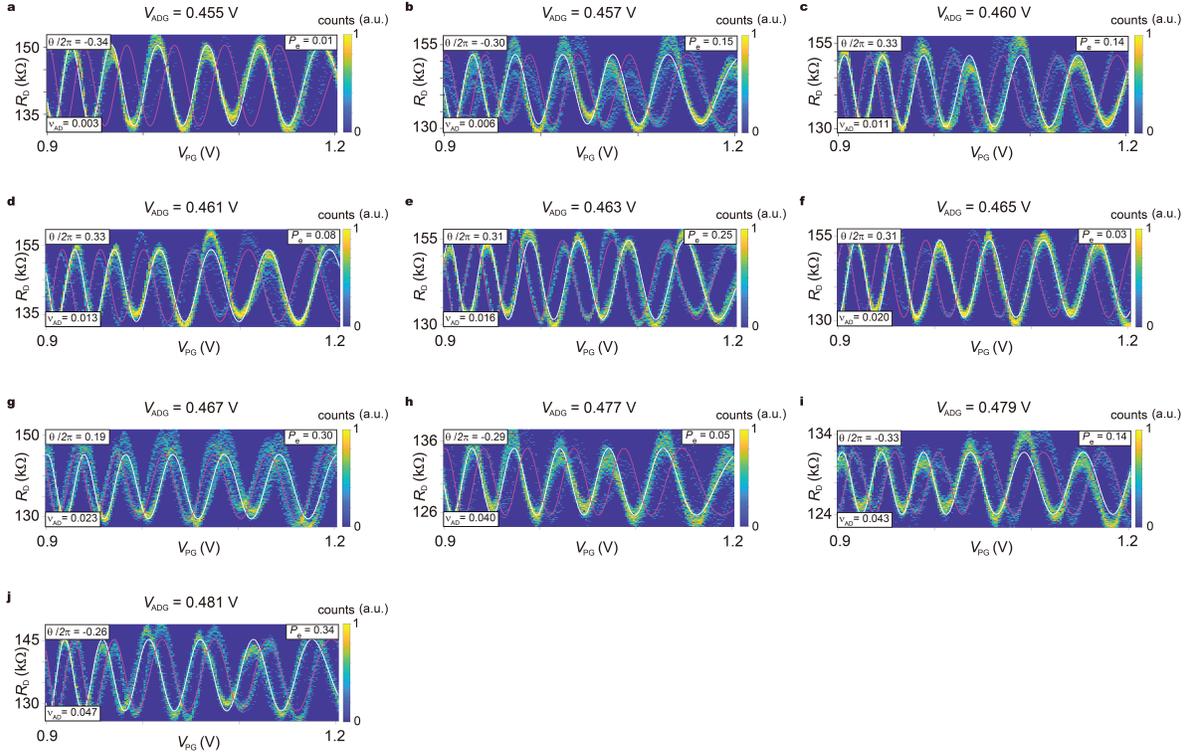

**Figure S9: Histograms of $R_D$ as a function of $V_{PG}$ for $\nu_{AD} > 0$.** (a-j) Histograms of $R_D$ at $\nu = -1/3$ as a function of $V_{PG}$, constructed from repeated measurements of $R_D$ under different $V_{ADG}$. **(a)** $\nu_{AD} = 0.003$. **(b)** $\nu_{AD} = 0.006$. **(c)** $\nu_{AD} = 0.011$. **(d)** $\nu_{AD} = 0.013$. **(e)** $\nu_{AD} = 0.016$. **(f)** $\nu_{AD} = 0.020$. **(g)** $\nu_{AD} = 0.023$. **(h)** $\nu_{AD} = 0.040$. **(i)** $\nu_{AD} = 0.043$. **(j)** $\nu_{AD} = 0.047$. $V_{ADG}$ values corresponding to each $\nu_{AD}$ are indicated above each panel. The white and red curves are cosine fits to the histogram peaks, clearly demonstrating two different phases of oscillating $R_D$.

13

## SI8 Histograms of $R_D$ at $\nu = -1/2$ under 8 T

Figures S10 and S11 show two-dimensional histograms of $R_D$ constructed at each $V_{PG}$ using the same method described in SI7. The phase difference between the two fitted curves is approximately $\pi$ for $\nu_{AD} \approx -1/2$, whereas a smaller phase difference is observed for $\nu_{AD} \approx 0$.

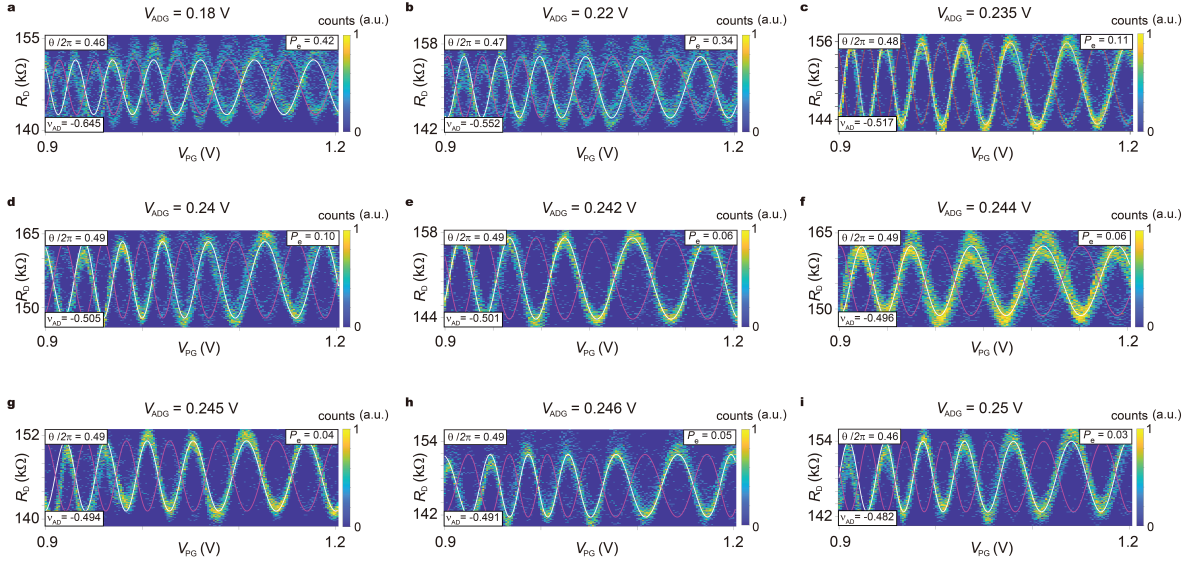

**Figure S10: Histograms of $R_D$ as a function of $V_{PG}$ for $\nu_{AD} \approx -0.5$.** (a-i) Histograms of $R_D$ at $\nu = -1/2$ as a function of $V_{PG}$, constructed from repeated measurements of $R_D$ under different $V_{ADG}$. **(a)** $\nu_{AD} = -0.654$. **(b)** $\nu_{AD} = -0.552$. **(c)** $\nu_{AD} = -0.517$. **(d)** $\nu_{AD} = -0.505$. **(e)** $\nu_{AD} = -0.501$. **(f)** $\nu_{AD} = -0.496$. **(g)** $\nu_{AD} = -0.494$. **(h)** $\nu_{AD} = -0.491$. **(i)** $\nu_{AD} = -0.482$. $V_{ADG}$ values corresponding to each $\nu_{AD}$ are indicated above each panel. The white and red curves are cosine fits to the histogram peaks, clearly demonstrating two different phases of oscillating $R_D$.

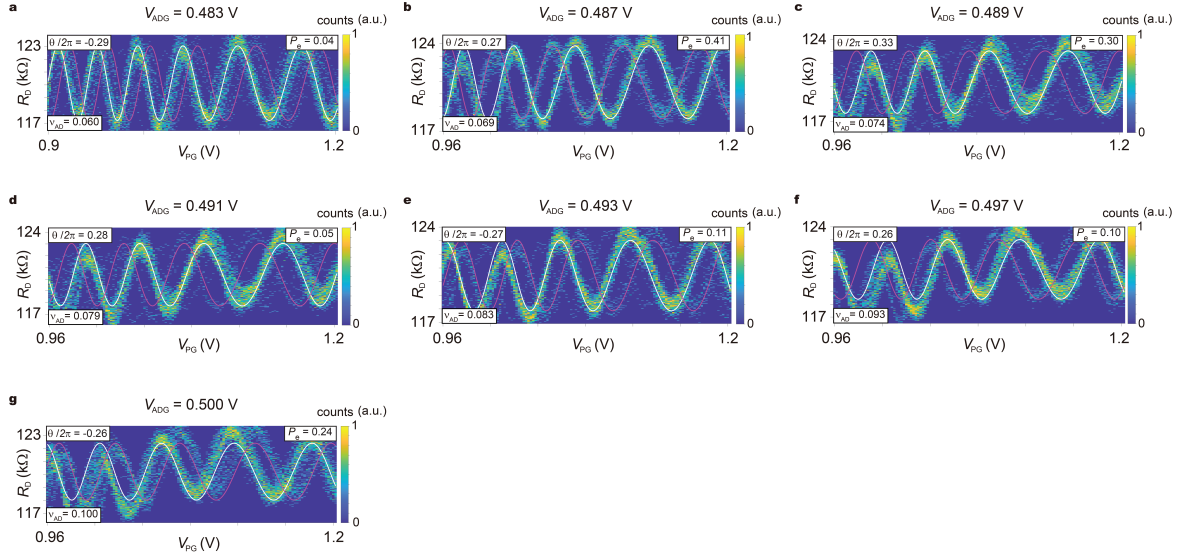

**Figure S11: Histograms of $R_\text{D}$ as a function of $V_\text{PG}$ for $\nu_\text{AD} \approx 0$.** **(a-g)** Histograms of $R_\text{D}$ at $\nu = -1/2$ as a function of $V_\text{PG}$, constructed from repeated measurements of $R_\text{D}$ under different $V_\text{ADG}$. **(a)** $\nu_\text{AD} = 0.060$. **(b)** $\nu_\text{AD} = 0.069$. **(c)** $\nu_\text{AD} = 0.074$. **(d)** $\nu_\text{AD} = 0.079$. **(e)** $\nu_\text{AD} = 0.083$. **(f)** $\nu_\text{AD} = 0.093$. **(g)** $\nu_\text{AD} = 0.100$. $V_\text{ADG}$ values corresponding to each $\nu_\text{AD}$ are indicated above each panel. The white and red curves are cosine fits to the histogram peaks, clearly demonstrating two different phases of oscillating $R_\text{D}$.

## SI9 Extraction of $P_{ex}$ from histogram populations

To quantify the relative population of the two interference phases, we analyzed the two-dimensional histograms of $R_D$ obtained from repeated measurements. Figure S12 shows this analysis applied to the histogram presented in Fig. 1f of the main text. In Fig. S12a, a reference value $R_{D0}$ was defined between the two histogram ridges to separate the two populations corresponding to the different interference phases. For each $V_{PG}$, the counts above and below $R_{D0}$ were summed to obtain the populations of the upper and lower branches, respectively. In Fig. S12b, the corresponding ratios were calculated by normalizing each population by the total counts, yielding $P_{upper}$ for the upper branch and $P_{lower}$ for the lower branch. To extract a representative value of the excitation probability $P_{ex}$, we identified plateau regions where the ratios vary weakly as a function of $V_{PG}$. The average values of the ratios within the plateau regions were then used to estimate $P_{ex}$, while the standard deviation provides an estimate of the uncertainty.

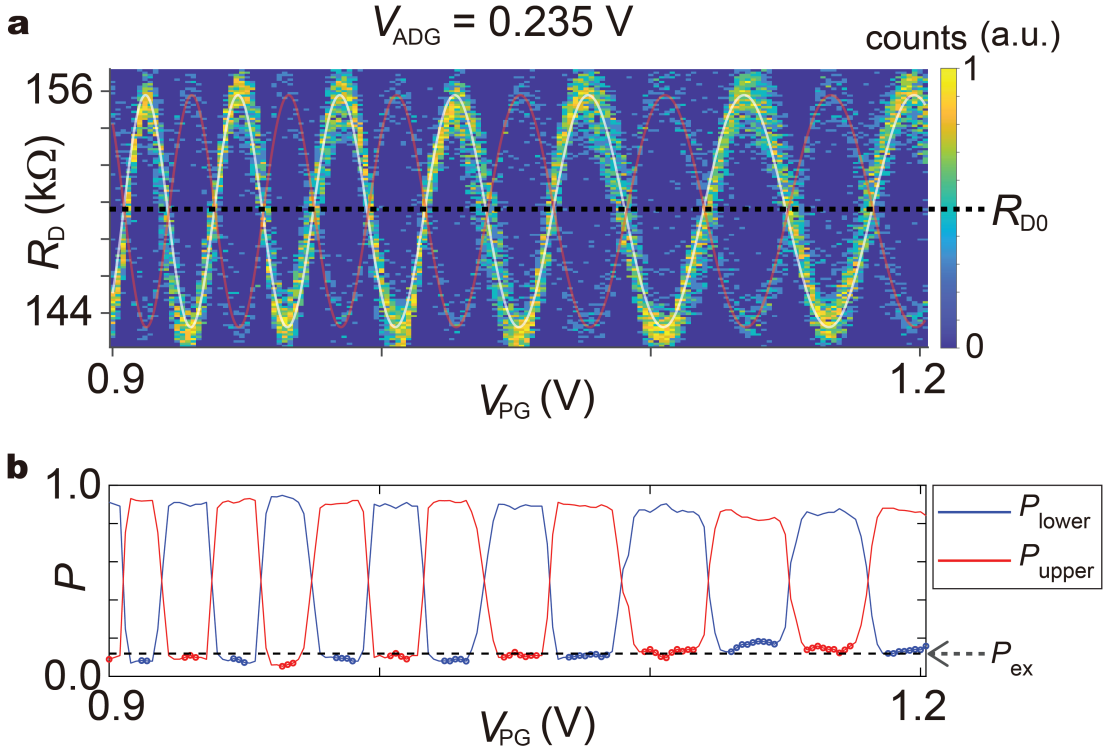

**Figure S12: Extraction of $P_{ex}$ from histogram populations.** (a) Histograms of $R_D$ at $\nu = -1/2$ as a function of $V_{PG}$, constructed from repeated measurements of $R_D$. The dashed line indicates the reference value $R_{D0}$ used to separate the two histogram ridges corresponding to different interference phases. (b) Populations of the two branches obtained by summing the histogram counts above and below $R_{D0}$, normalized by the total counts. The resulting ratios define $P_{upper}$ and $P_{lower}$. Plateau regions where the ratios remain approximately constant are used to extract the excitation probability $P_{ex}$ (gray dashed line).

## SI10 Quasiparticles and edge modes of paired quantum Hall states

Incompressible FQH states arise in a half-filled Landau level when composite fermions (CFs) form a paired superfluid. Such superfluids are classified by an integer topological index $\ell$ that corresponds to the number of chiral Majorana fermions $\psi$ at the boundary, i.e.,

$$\mathcal{L}_{\text{Majorana}} = i \sum_{k=1}^{|\ell|} \int_x [\psi_k(\partial_t - \text{sgn}(\ell)v\partial_x)\psi_k]. \tag{S1}$$

The integer $\ell$ also determines the number of Majorana zero modes in the core of a half-quantum ($\pi$) vortex, i.e., the analogue of an $hc/2e$ vortex in charged superfluids. Each of the chiral Majorana fermion theories in $\mathcal{L}_{\text{Majorana}}$ describes an Ising conformal field theory with three primary fields: The identity, the fermion $\psi_k$ with scaling dimension $\frac{1}{2}$ and the twist field $\sigma_k$ with dimension $\frac{1}{16}$.

In addition to the neutral Majorana fermions, the physical FQH state of electrons described by a CF superfluid hosts a bosonic charge mode at the boundary, described by

$$\mathcal{L}_{\text{charge}} = \frac{1}{2\pi} \int_x \partial_x \phi (\partial_t - v_c \partial_x)\phi. \tag{S2}$$

The charge mode has the same chirality as the Majorana fermions for positive $\ell$, which is the case for the Moore-Read Pfaffian ($\ell = 1$), but not for the anti-Pfaffian ($\ell = -3$).

Bulk quasiparticle excitations of the FQH fluid correspond to Bogoliubov quasiparticles and vortices of the CF superfluid. In particular, a $\pi$ vortex describes an $e/4$ excitation, which hosts an unpaired Majorana zero mode when $\ell$ is odd. On the boundary, this quasiparticle corresponds to $\Psi_{e/4} \sim \prod_k \sigma_k e^{i\phi/2}$. The Bogoliubov quasiparticles describe charge-neutral fermion excitations of the FQH fluid that carry odd fermion parity $\Psi_0^{\text{odd}} \sim \psi_k$. Finally, $2\pi$ vortices describe $e/2$ quasiparticles and come in two flavors differing by their fermion parity, $\Psi_{e/2}^{\text{even}} \sim e^{i\phi}$ and $\Psi_{e/2}^{\text{odd}} \sim e^{i\phi}\psi_k$. The scaling dimensions of these operators are

$$[\Psi_{e/4}] = \frac{1 + |\ell|}{16}, \qquad [\Psi_0^{\text{odd}}] = \frac{1}{2}, \qquad [\Psi_{e/2}^{\text{even}}] = \frac{1}{4}, \qquad [\Psi_{e/2}^{\text{odd}}] = \frac{3}{4}. \tag{S3}$$

The braiding phases of all quasiparticles are obtained by adding the contributions from the bosonic and Ising sectors. In the bosonic sector, encircling $e^{in\phi}$ with $e^{im\phi}$ yields $\theta_{\text{braid}}(e^{in\phi}, e^{im\phi}) = mn\pi$, which can be viewed as the Aharonov-Bohm phase resulting from a charge $me/2$ encircling a flux $2\pi n$. In the Ising sector, the full braiding of two fermions is trivial, while braiding $\psi_k$ around $\sigma_k$ yields $\theta_{\text{braid}}^{\sigma\psi} = \pi$, reflecting the sign change of a fermion encircling a half-quantum vortex. Consequently, braiding of two $e/2$ quasiparticles yields $\theta_{\text{braid}} = \pi$ independently of their parity, while

$$\theta_{\text{braid}}(\Psi_{e/2}^{\text{even}}, \Psi_{e/4}) = \frac{\pi}{2} \qquad \theta_{\text{braid}}(\Psi_{e/2}^{\text{odd}}, \Psi_{e/4}) = -\frac{\pi}{2}. \tag{S4}$$

For a comprehensive discussion of Abelian and non-Abelian paired states, see Refs. 5 and 6.



\end{document}